%% file: main.tex
\documentclass[10pt,journal,compsoc]{IEEEtran}
\usepackage[T1]{fontenc}
\usepackage[utf8]{inputenc}\DeclareUnicodeCharacter{2212}{-}
\usepackage[table]{xcolor}
\usepackage{pgfplots}
\usepackage{graphicx}
\usepackage{adjustbox}
\usepackage{amsmath}
\usepackage{makecell}
\usepackage{truncate}
\usepackage{bold-extra}
\usepackage[natbib=true,backend=biber,style=ieee, minnames=1, maxcitenames=2]{biblatex}
\usepackage[binary-units=true]{siunitx}
\usepackage[flushleft]{threeparttable}
\usepackage{stackengine}
\usepackage{paralist}
\usepackage{cellspace}
\usepackage{hhline}
\usepackage{tcolorbox}
\usepackage{placeins}



\usepackage{layouts}
\usepackage{booktabs}
\usepackage[boldmath]{numprint}
\usepackage{dcolumn}
\usepackage{multirow}
\usepackage[hidelinks]{hyperref}
\usepackage{rotating}
\usepackage{wasysym}
\usepackage{pifont}%
\usepackage{cleveref}
\usepackage{colortbl}
\usepackage{tabularx}
\usepackage{soul}
\usepackage{csquotes}
\usepackage{titlesec}

\titleformat{\paragraph}[runin]{\normalfont\normalsize\bfseries}{\theparagraph}{1em}{}[. \mbox{}]
\titlespacing*{\paragraph}{0pt}{1ex plus .5ex minus .2ex}{0pt}

\newcommand{\cmark}{\ding{51}}%
\newcommand{\xmark}{\ding{55}}%

\makeatletter
\let\MYcaption\@makecaption
\makeatother
\usepackage[font=footnotesize]{subcaption}
\makeatletter
\let\@makecaption\MYcaption
\makeatother

\pgfplotsset{compat=1.17}

\newcolumntype{d}[1]{D{.}{.}{#1} }
\newcommand{\bftab}{\fontseries{b}\selectfont}

\newcommand{\dtoprule}{\specialrule{1pt}{0pt}{0.6pt}%
            \specialrule{0.3pt}{0pt}{0pt}%
            }
\newcommand{\dbottomrule}{\specialrule{0.3pt}{0pt}{0.6pt}%
            \specialrule{1pt}{0pt}{\belowrulesep}%
            }
\newcommand{\dmidrule}{\specialrule{0.3pt}{0pt}{0.6pt}%
            \specialrule{0.3pt}{0pt}{0pt}%
            }

\include{defs}

\def\arminer{\textsc{AR-Miner}}

\newcommand*\conclusion[1]{
\begin{tcolorbox}[left=0mm,right=0mm,top=0.1cm, bottom=0.1cm, boxrule=0.25mm,colback=gray!5!white, enlarge top by=0.2cm, enlarge bottom by=0.2cm]
#1
\end{tcolorbox}
}

\renewcommand{\cite}{\citep}

\begin{document}

\title{Making the most of small Software Engineering datasets with modern machine learning}

\author{Julian~Aron~Prenner, %
        Romain~Robbes%
        
\thanks{Julian Aron Prenner and Romain Robbes are with the Faculty of Computer Science, Free University of Bozen-Bolzano, Italy}}

\IEEEtitleabstractindextext{%
\begin{abstract}
This  paper provides a starting point for Software Engineering (SE) researchers and practitioners faced with the problem of training machine learning models on small datasets. Due to the high costs associated with labeling data, in Software Engineering, there exist many small (< \numprint{1000} samples) and medium-sized (< \numprint{100000} samples) datasets. While deep learning has set the state of the art in many machine learning tasks, it is only recently that it has proven effective on small-sized datasets, primarily thanks to pre-training, a semi-supervised learning technique that leverages abundant unlabelled data alongside scarce labelled data.

In this work, we evaluate pre-trained Transformer models on a selection of 13 smaller datasets from the SE literature, covering both, source code and natural language. %
Our results suggest that pre-trained Transformers are competitive and in some cases superior to previous models, especially for tasks involving natural language; whereas for source code tasks, in particular for very small datasets, traditional machine learning methods often has the edge. 

In addition, we experiment with several techniques that ought to aid training on small datasets, including active learning, data augmentation, soft labels, self-training and intermediate-task fine-tuning, and issue recommendations on when they are effective. We also release all the data, scripts, and most importantly pre-trained models for the community to reuse on their own datasets.

\end{abstract}

\begin{IEEEkeywords}
Small Datasets, Transformer, BERT, RoBERTA, Pre-training, Fine-Tuning, Data Augmentation, Back Translation, Soft Labels, Active Learning.
\end{IEEEkeywords}}

\maketitle

\IEEEraisesectionheading{\section{Introduction}}
\IEEEPARstart{S}{mall} datasets are commonplace for many Software Engineering problems. While the creation of a labelled dataset is always a significant undertaking, this is even more the case for Software Engineering. In many cases, significant expert knowledge is required to label Software Engineering data, making it difficult to use crowd-sourcing techniques, as is often done in other fields such as in computer vision \cite{deng2009imagenet}. Moreover, some labelling tasks involve detailed (text or source code) understanding, making the labelling of a single example time consuming. Dataset size may be further reduced by the need to label the same examples multiple times and to compute inter-rater agreement. Due to all these factors, it is thus not uncommon for hand-labelled Software Engineering datasets to number only a few thousands or even hundreds of samples. For instance, the 13 datasets used in this work (described in Section \ref{sec:datasets}) range from 200 to 62,275 samples, with three datasets having more than 5,000 samples, and four having less than 1,000.

Historically, small Software Engineering datasets were used with traditional machine learning algorithms, such as Support Vector Machines (SVMs), Logistic Regression or Random Forests, often combined
with manual feature engineering. In recent years, early experiments with deep learning architectures \cite{codesmell, howfarcanwego, satd-cnn, satd-lstm}, such as Convolutional Neural Networks (CNNs) and Recurrent Neural Networks (RNSs), showed mixed results, suggesting that for many tasks where training data is scarce, deep learning does not provide a clear benefit, especially in light of its considerably higher computational costs. 

Whether deep learning is in fact not well suited for these small datasets, and if so, for which kind of tasks and dataset sizes is the central question of this paper. The motivation to take a second look at this problem is the recent advent of semi-supervised learning \cite{zhu2009introduction}. Semi-supervised learning is a machine learning paradigm in which both labelled data and unlabelled data are leveraged in the learning process, with the latter being much cheaper to acquire. While prevalent in computer vision, it is only since 2018 that semi-supervised learning has become viable in the NLP domain \cite{peters2018deep, howard2018universal}, in the form of pre-training. Since 2019, pre-trained Transformer-based models such as BERT~\cite{bert} or RoBERTa~\cite{roberta} have set many records in NLP and related fields, and considerably improved the state of the art on important benchmarks such as GLUE~\cite{glue} and SQuAD~\cite{squad}, in which there are small datasets. In addition to pre-training, several additional techniques have the potential to benefit small datasets, including Domain-specific Fine-Tuning, Intermediate-Task fine-tuning, Active Learning, Self-Training, Data Augmentation, and Soft Labels. Thus, a better understanding of when to combine these techniques is necessary. We provide background on the Transformer Architecture, Semi-supervised learning via pre-training and other techniques in Section \ref{sec:background}.

While previous work showed promising results in applying pre-training to SE problems \cite{robbes2019leveraging, mahadi2020cross, howfarcantransformersgo, norbert, bert4sentise}, this work examines this phenomenon in more depth, by applying the pre-training paradigm on thirteen different small and medium-sized Software Engineering datasets selected from the literature. These datasets span natural language, source code, and source code comments, in a variety of domains (several sentiment analysis tasks, several app review classification tasks, technical debt detection, comment classification, code comment coherence, code smell detection, code readability, code complexity). In addition, and unlike previous work, we also investigate the impact of the additional techniques mentioned above, when they are relevant. Section \ref{sec:method} presents methodological details such as pre-processing, baselines, and training, testing and validation modalities, for all the scenarios we consider. This section also presents the pre-trained and fine-tuned models we use in this work, including \stackobertflowshort{}, a Transformer model pre-trained on 26 million Stack Overflow comments. 

Section \ref{sec:results} presents the results of the paper, answering the following research questions:
\begin{itemize}
\item RQ1: For which domains and tasks does the pre-training paradigm outperform the baselines? We find that pre-training is effective for tasks working on natural language and source code comments, but is not as effective for tasks working on source code yet.
\item RQ2: Which additional techniques are effective, and if so in which circumstances? We find that some techniques, such as domain-specific pre-training, and data augmentation are effective in some (but not all) settings, while we find limited evidence for the effectiveness of others, such as active learning.
\end{itemize}

Finally, we close the paper by documenting the limitations of our study, and the opportunities for additional studies in Section \ref{sec:limit}. We conclude the work in Section \ref{sec:conc}, summarizing initial recommendations on the effectiveness of pre-training and the additional techniques. Additional material can be found in three appendices: Appendix \ref{appendix:datasets} provides additional information on datasets; Appendix \ref{appendix:results} provides results; last but not least, Appendix \ref{appendix:implementation} provides instructions on how to access the data, scripts, and pre-trained models we used in our experiments. These models can be fine-tuned for a wide range of tasks relating to software artifacts; we hope that they will prove useful to other researchers in the field.

\section{Datasets and Related Work}\label{sec:datasets}

We selected thirteen datasets introduced in the Software Engineering literature in recent years, aiming for both variety in terms of artifacts, dataset size, and classification tasks
\cite{senti4sd, twodssenti, howfarcanwego, arminer, review-classification, satd-orig, comment-classification, corazza-ds, codesmell, complex, scalabrino2016}. Another selection factor was the availability of a comparable baseline or a way to reproduce the initial experiment.
Seven datasets involve natural language, two code comments, and the remaining four source code (one with comments). They vary from 341 to 62,275 examples, and from 2 to 16 classes.
The datasets cover nine different tasks: 
\begin{inparaenum}[a)]
\item sentiment classification of software artifacts, such as Stack Overflow comments, app and code reviews
\item detection of informative app reviews
\item classification of app reviews
\item detection of self-admitted technical debt through code comments
\item classification of code comments
\item prediction of code-comment coherence
\item detection of linguistic code smells
\item prediction of code runtime complexity and
\item prediction of code readability.
\end{inparaenum} 

Next, we discuss each of these tasks in greater detail. For a more
concise overview refer to Table~\ref{tab:ds-overview}.

\newcommand{\noimprov}{\xmark}%
\newcommand{\lowimprov}{\stackengine{0pt}{\cmark}{\xmark}{O}{c}{F}{F}{L}}%
\newcommand{\improv}{\cmark}%

\definecolor{Gray}{gray}{0.95}
\newcolumntype{B}{>{\columncolor{Gray}}c}

\setlength\cellspacetoplimit{3pt}
\setlength\cellspacebottomlimit{3pt}

\begin{table}
  \footnotesize
  \makegapedcells
  \begin{center}
  {\def\arraystretch{1}\tabcolsep=2pt\setcellgapes{3pt}
  \begin{tabular}{crccc}
    \toprule
    \textbf{Name}&\textbf{Size}&\textbf{\#\,Cl.}&\textbf{Type}&\textbf{Usage}\\
    \midrule
    \makecell[c]{Sentiment Classification\\(Stack Overflow)~\cite{senti4sd}} & \numprint{4423} & 3 & \makecell{Natural\\language} & \makecell{Train, Test,\\Valid.} \\
    \makecell[c]{Sentiment Classification\\(Stack Overflow)~\cite{twodssenti}} &  \numprint{1500} & 3 & \makecell{Natural\\language} & Test \\
    \makecell[c]{Sentiment Classification\\(JIRA Issues)~\cite{howfarcanwego}} &  \numprint{926}  & 2 & \makecell{Natural\\language} & Test \\
    \makecell[c]{Sentiment Classification\\(App Reviews)~\cite{twodssenti}} &  \numprint{341}  & 3 & \makecell{Natural\\language} & Test \\
    \makecell[c]{Informative App Review\\Detection~\cite{arminer}} & \numprint{12000} & 2 & \makecell{Natural\\language} & \makecell{Train, Test,\\Valid.} \\
    \makecell[c]{App Review\\Classification~\cite{review-classification}} & \numprint{3691} & 4 & \makecell{Natural\\language} & Train, Valid. \\
    \makecell[c]{App Review\\Classification~\cite{clap}} & \numprint{3000} & 7 & \makecell{Natural\\language} & Train, Valid. \\    
    \midrule
    \makecell[c]{Self-Admitted Technical\\Debt Detection~\cite{satd-orig}} & \numprint{62275}      & 2 & Comments & \makecell{Train, Test,\\Valid.} \\
    \makecell[c]{Comment Classification\\\cite{comment-classification}} & \numprint{11232} & 16 & Comments & Train, Valid. \\
    \midrule    
    \makecell[c]{Code-Comment Coherence\\Prediction~\cite{corazza-ds}} & \numprint{2881} & 2 & \makecell{Code w/\\Comments} & \makecell{Train, Test,\\Valid.} \\
    \makecell[c]{Linguistic Smell Detection~\cite{codesmell}} & \numprint{1753} & 2 & Code & Train, Valid. \\
    \makecell[c]{Code Runtime Complexity\\Classification~\cite{complex}} & \numprint{933} & 5 & Code & Train, Valid. \\
    \makecell[c]{Code Readability\\Prediction~\cite{scalabrino2016}} & \numprint{200} & 2 & Code & Train, Valid. \\
    \bottomrule
  \end{tabular}
  }
  \end{center}
  \caption{Datasets considered in this work, along with their size, number of classes and usage.} 
  \label{tab:ds-overview}
\end{table}

\subsection{Natural Language Datasets}

These datasets contain mainly natural language, but may occasionally contain some source code identifiers. They are the closest to the original setting for models pre-trained on a generic English corpus, although they come from very specific domains.

\subsubsection*{Sentiment Classification}

In sentiment classification, a model assigns a sentiment class (e.g., one of \textit{positive}, \textit{negative}, \textit{neutral}) to a sentence or short piece of text; in our case the text's domain is related to software development.

\AtNextCitekey{\defcounter{maxnames}{1}\defcounter{minnames}{1}}
The dataset compiled by \citeauthor{senti4sd} (Senti4SD) contains \numprint{3097} training and \numprint{1326} test samples each labeled
as either \textit{positive}, \textit{negative} or \textit{neutral}; all samples were extracted from questions, answers and comments of Stack Overflow posts.
Along with their dataset, the authors also released individual rater annotations, i.e., three rater labels per sample, from which the final
labels were obtained by applying a majority vote rule. The dataset is used to evaluate an SVM classifier using word embeddings trained on a Stack Overflow corpus as features;
we include this SVM as a baseline (see Table \ref{tab:senti-results}). 

\AtNextCitekey{\defcounter{maxnames}{1}\defcounter{minnames}{1}}
\citeauthor{howfarcanwego} created a sentiment classification datasets consisting of 1,500 sentences (from 178 text fragments) extracted from Stack Overflow discussions. They also  adapted two previous datasets of 636 Jira issues (926 sentences), and 130 app reviews (341 sentences) \citep{sentijira, release-planning-mobile}. 
The JIRA issues dataset has only two sentiment classes (\textit{positive} and \textit{negative}) while the other two have an additional \textit{neutral} class. All three datasets were used in a study of several sentiment analysis tools, and a novel model introduced in the same work \citep{howfarcanwego};
we include all of them as baselines in our comparison  (Table \ref{tab:senti-results}).
To allow for a comparison with these baselines, all three datasets are \textit{only used as test sets} in this work.

\subsubsection*{Informative App Review Detection}
An app review is considered informative if it contains valuable information for the application developer,
such as feature suggestions or bug reports. The dataset presented by \citet{arminer} contains \numprint{12000} app reviews belonging to four popular mobile apps from the Google Play Store. The dataset is partitioned into predefined test (2000 samples per app) and train sets (1000 samples per app). Three raters annotated each review as either \textit{informative} or \textit{non-informative}, with a majority vote to determine the final label.  This work also presented \arminer{}, a tool based on an expectation maximization with Naive Bayes (EMNB) classifier to detect such informative app reviews; this tool is our baseline.

\subsubsection*{App Review Classification}

\citet{review-classification} compiled a dataset of \numprint{4400} reviews
crawled from Apple's App Store and Google Play. Each review was categorized by two raters into one of four classes (\textit{bug report}, \textit{feature request},
\textit{user experience} or \textit{rating}); reviews with rater disagreement were discarded.
The authors experimented with various types of classifiers, finding that an ensemble of binary classifiers performs considerably better than a single multiclass classifier.
We only study multiclass classification, for comparability reasons, use the author's multiclass classifier as a baseline. \citet{release-planning-mobile, clap} introduce CLAP, a tool for automatic classification and clustering of app reviews that is evaluated on a dataset of \numprint{3000} app reviews created by the same authors. We use the CLAP dataset in a data augmentation experiment.

\subsubsection*{Related Work}
There is a vast literature on sentiment analysis and classification. For a general overview see e.g.,~\citet{MANTYLA201816}.
\citet{howfarcantransformersgo} provide an in-depth comparison of pre-trained Transformers with six different sentiment analysis tools, including both, tools for general sentiment analysis and tools specifically targeted towards Software Engineering.
\AtNextCitekey{\defcounter{maxnames}{1}\defcounter{minnames}{1}}
The study uses some datasets that we also use (e.g., from \citet{howfarcanwego}), they defined custom training and testing sets while we used them solely for testing, which makes comparisons difficult. While they explored additional Transformer architectures (XLNet~\cite{xlnet} and ALBERT~\cite{albert}), their study was limited to only fine-tuning: they did not investigate the use of task-specific pre-training, nor any other of the additional methods that we study. %
\citet{senticr}, introduced SentiCR, a sentiment analysis tool that uses of Part Of Speech (POS) tags and Gradient Boosting Trees, while \citet{sentimoji} present SEntiMoji, a model that leverages emojis to improve SE sentiment classification.

\citet{review-classification-active-learn} investigated the application of \textit{active learning} for app review analysis, using the dataset
\AtNextCitekey{\defcounter{maxnames}{1}\defcounter{minnames}{1}} by \citeauthor{review-classification}. They employ traditional machine learning
algorithms (naive Bayes, logistic regression, and SVM). A similar experiment, carried out on the same dataset can also be found in this work.

\subsection{Datasets of Code Comments}

Source code comments somewhat differ from natural language: they may often contain source code identifiers, code annotations, and specific idioms common in source code documentation.

\subsubsection*{Self-Admitted Technical Debt Detection}
\textit{Self-admitted} technical debt (SATD) is technical debt known to and acknowledged by the author. It is often expressed in code comments with
a short description of a flaw or shortcoming and sometimes, but not always, marked with specific keywords, such as \texttt{FIXME}, \texttt{TODO} or \texttt{HACK}.
Detection of such comments can be useful to assess software quality, aid decision-making or direct further development.

\AtNextCitekey{\defcounter{maxnames}{1}\defcounter{minnames}{1}}
The dataset by \citeauthor{satd-orig} contains SATD comments extracted from 10 prominent Java projects (for more details, see Table \ref{tab:satd-ds} in Appendix~\ref{appendix:datasets}).
With over \numprint{60000} samples, it is by far the largest dataset used in this work.
Each dataset sample is assigned to one of five SATD categories informed by an established ontology of technical debt~\citep{satd-ontology} (\textit{design debt}, \textit{requirement
debt}, \textit{defect debt}, \textit{documentation debt} or \textit{test debt}) or labeled as not containing any SATD at all.
In this work, we concentrate on the binary version of the problem (detecting presence of SATD), for two reasons: (1) because there are large class imbalances  
(i.e., the document debt class makes up less than 0.1\% of the total data), and (2) to compare performance to binary classifiers from previous work. In addition to introducing this dataset, the authors also perform various SATD detection and classification experiments using traditional NLP and machine learning methods;
we use their SATD detection model as a baseline. We also include the CNN from \citet{satd-cnn} as a second baseline.

\subsubsection*{Comment Classification}

\citet{comment-classification} released a dataset of over \numprint{11000} comments from open-source Java projects classified according to
a taxonomy of 16 different comment categories. They evaluate their dataset on a multinomial Naive Bayes classifier
which serves as our baseline for this task.

\subsubsection*{Related Work} 

For a more general survey on self-admitted technical debt see e.g., \citet{satd-survey}.
\citet{satd-lstm} use a Long short-term memory network (LSTM) to 
\AtNextCitekey{\defcounter{maxnames}{1}\defcounter{minnames}{1}}
\textit{classify} SATD, also making use of the dataset by \citeauthor{satd-orig}; since we confined ourselves to SATD \textit{detection} this work was not included. A text-mining based approach
to SATD detection can be found in \citep{satd-tm1, satd-tm2}; in this work, unfortunately, only a subset of the dataset was used and results are thus not comparable. In a closely related comment classification work, \citet{comment-classification-mobile} focus on comment classification in mobile applications.

\subsection{Datasets of Source Code}

These datasets significantly differ from natural language: on the one hand, code has a very specific and unambiguous syntax and are much more repetitive than natural language \cite{hindle2016naturalness}; on the other hand, code has very complex semantics, and has many identifiers, leading to vocabulary issues \cite{karampatsis2020big}.

\subsubsection*{Code-Comment Coherence Prediction}
\citet{corazza-ds, corazza, corazza-we} examine the concept of code-comment coherence, i.e., the \enquote{relatedness} of a method's code and its lead comment.
The authors introduce a dataset of \numprint{2883} \textit{Java} methods along with their leading comment and a binary label indicating whether
coherence exists between the two or not~\citep{corazza-ds}.In follow-up work, the authors trained an SVM classifier on their dataset using features based on tf-idf~\citep{corazza} and later word embedings~\citep{corazza-we}.
We include both of these models as baselines.

\subsubsection*{Linguistic Smell Detection}

\citet{codesmell} examine the automatic detection of \textit{linguistic} code smells (also known as linguistic antipatterns),
that is, code smells emerging from the use of misleading identifier names or the violation of common naming conventions.
Examples of this are variable names as if they were lists or arrays when in fact they have a scalar type,
or getter methods with side effects. 

They labeled a dataset of roughly \numprint{1700} code
snippets, following a taxonomy of linguistic smells~\citep{codesmell-taxonomy}, which comprises 18 different types of linguistic antipatterns.
They then trained a number of models on this dataset and compared their performances. These models include CNNs in various configurations, SVMs with different kernels and 
a Random Forest classifier. All models are binary, that is, they only determine whether a given sample is \enquote{smelly} or not and do distinguish between different types of antipatterns.
Interestingly, the authors found that a thoroughly tuned SVM model outstrips the CNN in all its configurations. This not only in terms of performance
metrics but also in terms of memory consumption, training time and ease of use. Their best-performing models were selected as baselines.

\subsubsection*{Code Runtime Complexity Classification}

\citet{complex} investigate the use of machine learning to automatically predict the code runtime complexity class (e.g., $\mathcal{O}(n^2)$) of short
programs. To this end, they collected \numprint{933} Java implementations of various algorithms from a competitive programming platform and annotated each
 with the corresponding complexity class (i.e., one of $\mathcal{O}(1)$, $\mathcal{O}(\log{}n)$, $\mathcal{O}(n)$, $\mathcal{O}(n\log{}n)$, $\mathcal{O}(n^2)$).
They experiment with various traditional machine learning approaches, training, such as Random Forests and SVMs on manually engineered features (such as numbers of loops, numbers of variables etc.) and code embeddings obtained from the programs' abstract syntax tree through graph2vec~\citep{graph2vec}.

\subsubsection*{Code Readability Prediction}
What constitutes readable code and what does not, seems to be largely a matter of personal taste. Notwithstanding this, research by \citet{buse} suggests that code readability
can, at least in part, be measured objectively.

A relatively small number of papers \cite{buse, posnett, dorn, scalabrino2016, scalabrino2018} examine models for automatic code readability estimation.
Most recently, \citet{scalabrino2016, scalabrino2018} compiled a dataset %
by letting 30 Computer Science students rate the readability of 200 methods, previously selected from well-known Java projects.
Each method received 9 readability ratings: these ratings were then averaged and compared against
a threshold value to assign a single binary readability label.
Further, they developed a logistic regression model for code readability estimation by combining structural readability features
proposed in \citet{buse} and \citet{dorn} with novel textual features. 
We base all of our experiments on above dataset; previous datasets by \citet{buse} and \citet{dorn} were not included due
to lack of baselines suitable for comparison.

\subsubsection*{Related Work}
\citet{code-comment-understanding} analyzed code-comment coherence by means of a Bi-LSTM model which was evaluated also on the above dataset. However, because a different
evaluation methodology was used, a direct comparison was not possible. \citet{codesmell-taxonomy}, in addition to introducing the already mentioned code smell taxonomy, provide an exhaustive treatment of this subject, including also an in-depth empirical study 
of how developers perceive such smells. A system to detect linguistic code smells in infrastructure as code scripts was developed by \citet{codesmell-deepiac}.
There exists further literature on smell detection in a broader context: for instance, \citet{codesmell-fontana} presented a machine learning approach to code smell detection whose results, however, were called into question in a later replication study~\citep{codesmell-arewethereyet}. More recently, \citet{codesmell-deepl} used deep learning models such as CNNs and LSTM networks to detect
code smells in C\# and Java code.
Also related, is work done by \citet{codesmell-severity}, who experiment with machine learning models for code smell \textit{severity} prediction, which can be considered an extension of the
simpler detection task.

\section{Background}\label{sec:background}

This section provides an overview of the various machine learning techniques that we investigate in order to evaluate their effectiveness starting with the pre-training paradigm, then covering self-training, data augmentation, active learning, and soft labels. We also highlight the uses of pre-training in software engineering. 

\subsection{Pre-training with The Transformer and BERT}

\subsubsection*{The Transformer architecture} 
Transformer~\citep{transformer} networks are a relatively recent architecture, particularly popular in the domain of natural language processing (NLP). They have replaced Long short-term memory (LSTM) networks as the prevailing architecture for text-based data. Transformers are based on attention, a mechanism previously used in LSTM networks to align the information flow between the encoder and decoder part of the network~\citep{lstm-attention}. Attention allows
a model to connect related parts of a sentence and form complex structures of interdependence between them. 
Unlike LSTMs, Transformer networks are not recurrent and instead have a fixed-size input window of tokens (typically 512 tokens): this allows for
more efficient training and avoids vanishing-gradient or long-dependency problems extant in recurrent architectures.
To summarize, each layer of the Transformer uses the attention mechanism to learn relationships between its inputs,
which, in the case of the first layer are the input tokens; when used for classification, a final fully-connected layer is used as output layer.

\subsubsection*{Pre-training via Language Modelling} 
BERT~\citep{bert} (Bi-Directional Encoder Representations from Transformers) is an extension of the Transformer architecture and comes with a specific \textit{semi-supervised learning} training regimen: BERT heavily relies on \textit{pre-training}, a form of \emph{unsupervised learning}, before being \textit{fine-tuned} on a downstream task in a classical \textit{supervised} fashion. 

During pre-training, BERT is trained on large amounts of unlabeled data via Mask Language Modelling (MLM). MLM is a prediction task where some of the input tokens are randomly replaced by blanks (\enquote{masked}) and the model is trained to predict the tokens behind these blanks, taking into account the textual context on both sides of the blank (see the BERT paper for more details on the pre-training itself~\cite{bert}). Intuitively, this general task is supposed to initialize the weights to a state in which certain general concepts and relationships useful for a large number of downstream tasks are already present: BERT learns a \textit{Representation} of the tokens. Unlike word embeddings \cite{mikolov2013distributed}, these  are contextual representations: they depend both on the token, and its surrounding tokens.

\AtNextCitekey{\defcounter{maxnames}{1}\defcounter{minnames}{1}}
Of note, earlier work also used Language Modelling as a pre-training task (ELMo and ULMFit \citep{peters2018deep, howard2018universal}) with LSTMs, and were used with some varying amount of success in Software Engineering \citep{robbes2019leveraging, mahadi2020cross}. BERT's pre-training is more efficient for two reasons: BERT's bidirectional architecture uses the context before and after the token, whereas LSTMs use only the context before the token; and BERT uses Byte-Pair Encoding (BPE) \cite{sennrich2016neural} to tokenise text in subwords rather than entire words, leading to better modelling of the vocabulary (see previous work by \citeauthor{karampatsis2020big} for an extended discussion of this aspect for source code \citep{karampatsis2020big}). 

RoBERTa~\citep{roberta} is a refinement of BERT, in particular relating to its pre-training regimen (e.g., RoBERTa uses a larger pre-training corpus, dynamic masking, and a variation of the pre-training task) and with only minor architectural changes (RoBERTa uses Byte-level BPE tokenization, rather than character-level BPE).

\subsubsection*{Fine-tuning} Both BERT and RoBERTa are hardly ever trained from scratch. Instead, starting from a pre-trained model with pre-initialized weights, the model weights are further \textit{fine-tuned} by
training on task-specific labeled data (called a downstream task). This involves replacing the last layer of the model (useful for the pre-training task), with a task-specific layer, and resuming training. The model can leverage the pre-trained representations to be able to learn the downstream task effectively, even with a limited amount of data, allowing BERT and RoBERTa to set the state of the art on NLP benchmarks, even on tasks with limited data (the GLUE benchmark \cite{glue} includes several task with less than 10,000 examples).

\subsubsection*{Impact of the Pre-training corpora} The standard BERT and RoBERTa models have both been pre-trained on a large English natural language corpus, with several models available in various sizes. There exist pre-trained BERT models for many other natural languages and even programming languages~\citep{codebert}. Intuitively, one would expect a generic pre-training corpus to be a \enquote{jack of all trades, master of none}, with a more specific pre-training corpus to be more suited for more specific domains (such as software engineering). There is evidence of this for word embeddings in Software Engineering \citep{efstathiou2018word}, but how much of an impact a domain-specific pre-training corpus has for a BERT or RoBERTa model is still an open question, which we investigate. Of note, the ULMFit approach \cite{howard2018universal} continues the pre-training task on the task-specific data (without using labels), before the actual fine-tuning, finding that it does improve performance.

\subsection{Additional Techniques} 

\subsubsection*{Intermediate-Task Fine-Tuning}
Intermediate-task fine-tuning (ITT), also known as two stage fine-tuning, STILTs~\citep{stilt},
or \textsc{Tanda}~\citep{tanda} is a technique whereby the model is fine-tuned twice (with labeled data):
first on an \textit{intermediate task}, a task different from but closely related to the target task, and finally on the actual target task (e.g., training for sentiment analysis on movies, before switching to sentiment analysis on books).
This is particularly attractive whenever only little data is available for the target task whilst
large amounts of data are available for a similar, possibly slightly simpler, but different intermediate task.
The idea is that the target task might benefit from \enquote{knowledge} that the model acquired
during intermediate-task training. \citet{intermediatetask} presents a survey on when this method
offers good prospects in NLP.

\subsubsection*{Self-Training}

\textit{Self-training} (also known as self-labelling or self-learning)~\citep{self-training1, self-training2}, is a very simple semi-supervised learning method. 
It can be explained as follows: A model is first trained on a (possibly too small) labeled dataset.
Next, this model is used to evaluate a number of additional \textit{unlabeled} samples. The model's predictions for these unlabeled samples are
then simply used as their gold labels. We now have additional labeled data, albeit noisier ones; after adding it to the original dataset we retrain the model.
Predictions can be filtered by confidence to reduce the probability of introducing noise into the training set.

\subsubsection*{Data Augmentation and Back-Translation}

\begin{figure}
\footnotesize
\begin{center}
\begin{tabularx}{\linewidth}{c X}
\toprule
\textbf{EN} & Leppie, that's great news! I look forward to trying IronScheme! \\
\midrule
\textbf{EN \textrightarrow DE} & Leppie, das sind großartige Neuigkeiten! Ich freue mich darauf, IronScheme auszuprobieren! \\
\textbf{DE \textrightarrow EN} & leppie, \ul{those are} great news! I \ul{am looking} forward to \ul{try out} IronScheme! \\
\midrule
\textbf{EN \textrightarrow FR} & Leppie, c'est une excellente nouvelle! J'ai hâte d'essayer IronScheme! \\
\textbf{FR \textrightarrow EN} & leppie, \ul{this is} great news! I \ul{can't wait to try} IronScheme! \\
\bottomrule
\end{tabularx}
\end{center}
\caption{Example of back-translation. The original English sentence is first translated to German and French, then
translated back into English; resulting variation underlined. Google Translate was used for the translation.} \label{fig:backtans-example}
\end{figure}

Data augmentation is a well-known technique to increase the amount of labeled data without any human labeling effort, which
is especially valuable in cases where training data is in short supply.
It works by adding slightly \textit{varied} copies of already existing, labeled samples to the dataset, assuming the variations do not affect the label. The technique was first used in computer vision, where data augmentations are easier to define, such as flipping images horizontally (a dog looking left instead of right), or cropping images randomly (a closeup of the dog's head should still be classified as a dog).
For text data, several such methods for augmentation have been proposed in recent years, among others: 
\begin{inparaenum}[a)]
\item replacing words with synonyms~\citep{aug-wordnet, aug-easy},
\item replacing, adding or deleting words randomly~\citep{aug-easy},
\item replacing words with the nearest neighbor in an embedding space~\citep{aug-glove, aug-word2vec},
\item replacing words with predictions from a masked language model such as BERT~\citep{aug-bert},
\item translating into an intermediate language and then back into the source language (back-translation)~\citep{aug-backtrans}.
\end{inparaenum}

Augmentation is typically applied at training time by simply adding 
the augmented samples to the training set and then proceeding as usual.
Alternatively, augmentation can also be carried out at test time by aggregating (e.g., averaging) the prediction for an original test sample
with the predictions for its augmented copies, thus obtaining potentially more stable or more accurate predictions. 

Figure \ref{fig:backtans-example} shows an example of augmentation through back-translation: a sample in the dataset (here an English sentence)
is translated into German and French, then back into English, causing slight variations. 

\subsubsection*{Active Learning}
The goal of active learning is to make the process of manual data labeling more efficient. Active learning avoids presenting samples to the rater that the model is likely to classify correctly and thus provide little new information.

Initially, a human rater labels a small number of samples, called the \textit{seed}.
There is also a second, larger set of yet unlabeled samples, called the \textit{pool}. A model is first trained on the seed.
In a next step, this model is used to select those samples from the pool that the model found most \enquote{difficult} to classify.
\enquote{Difficulty} is measured by means of a \textit{confidence} or \textit{acquisition function} which calculates a confidence score from the model's prediction. In classification, this is usually a distribution over the target classes and
acquisition functions are thus applied to class probabilities.
Selecting samples by confidence score is called \textit{confidence sampling}. %
The rater then labels the selected samples, which are then removed from the pool and added to the model's training set. This process is repeated until a satisfying number of samples have been labeled or the model reaches a particular target accuracy.
A possible problem with confidence sampling is that selected samples, albeit being difficult for the model, might all be very similar, reducing the efficiency of the process. %
Confidence sampling is often paired with \textit{diversity sampling}: selected samples are subsequently filtered for diversity, for instance using a clustering algorithm such as \textit{k}-means.
A common way to evaluate and compare active learning approaches is a \textit{simulation} with an already labeled dataset. See \citet{actlearn-survey} for an overview of variants and extensions of active learning. %

\subsubsection*{Soft Labels}

In classification, usually, every sample is associated with a single %
target class. %
For many machine learning algorithms, in particular for neural networks,
the target label of a sample is represented as a probability distribution over classes. While optimizations exist for handling the common single-label case, conceptually we can say that
the target label is denoted by a distribution vector which assigns probability one to the class it belongs to and probability zero to all other classes. Take, for example, a classification problem
where each sample belongs to either class $A$, $B$ or $C$. A sample will have the target vector ($A$: 0, $B$: 1, $C$: 0) if it belongs to class $B$ and 
($A$: 0, $B$: 0, $C$: 1) if it belongs to class $C$.

The term \textit{soft label} is used when this distribution vector is fuzzy, i.e., is not comprised of a single one and many zeros. Intuitively, this means
that a sample can belong to multiple classes, with a degree expressed by the class probabilities: a target distribution such as ($A$: 0.4, $B$: 0.6, $C$:0) belongs to both, class $B$ and class $A$. %
Most datasets do not come with soft-labels. In cases where each sample in a dataset was classified by multiple raters (which is common in order to compute inter-rater agreement), instead of using a majority vote, the rater's votes can be converted into a soft-label. Intuitively, an example in which raters disagree can be seen as more ambiguous. Providing this information to the model can help it differentiate between \enquote{easy} examples and \enquote{hard} examples.

\subsection{Uses of Pre-training and Transformers in Software Engineering}

In previous work, we investigated the usefulness of the pre-training paradigm (using the earlier ULMFit approach \citep{howard2018universal}), finding it promising in limited sentiment analysis experiments \citep{robbes2019leveraging}. \citet{mahadi2020cross} experimented with cross-dataset classification of design discussions, but had mixed results.

\citet{howfarcantransformersgo} provide a detailed study on the use of Transformers for sentiment analysis of Software Engineering artifacts, comparing
existing sentiment analysis tools with Transformer models (BERT, RoBERTa, XLNet). \citet{bert4sentise} pursue a similar avenue, training BERT on a newly compiled dataset of \numprint{4000} sentences from Stack Overflow discussions and comparing results with recurrent models.

\citet{norbert} present NoRBERT, a BERT model fine-tuned to classify functional and non-functional requirements
that achieves results competitive to state-of-the-art models on the PROMISE NFR dataset.

\citet{bert-arch-tactics} attempt to use a standard BERT (i.e., pre-trained on English natural language) for the detection of architectural tactics in Java code and 
report mixed results that lag behind state-of-the-art approaches in one case study.

\citet{intellicode-compose} introduce IntelliCode compose, a system for intelligent code completion based on the GPT-2 Transformer language model.

Finally, \citet{codebert} present CodeBERT, a RoBERTa-based Transformer that was trained on natural language and code (bimodal), allowing for code-related tasks that also involve natural language, such as code search or documentation generation.

\section{Methodology}\label{sec:method}
This section covers general aspects of the methodology, that apply to all the experiments. To ease readability, methodological details that refer to a specific technique (e.g., active learning) are described jointly with the results of this technique in Section~\ref{sec:results}.

\subsection{Pre-Trained models}\label{subsec:pretrain}

\subsubsection*{Off-the-shelf models}

\def\bertbase{BERT-base} 
\def\bertlarge{BERT-large}
\def\distillroberta{DistillRoBERTa}
\def\bertreviews{BERT-reviews}
\def\bertcomments{BERT-comments}
\def\bertSOM{BERT-SO-1M}
\def\bertSOMM{BERT-SO-2M}
\def\bertSOML{BERT-SO-1M-large}

We use several \enquote{off-the-shelf} pre-trained model, which were trained on a corpus of generic English text. 
\begin{itemize}
    \item \bertbase{}, a 12 layer Transformer model (~110 million parameters), pre-trained on a 3.3 billion words from books and Wikipedia \cite{bert}. 
    \item \bertlarge{},  a 24 layer Transformer model (~340 million parameters), pre-trained on the same corpus \cite{bert}.
    \item \distillroberta{}, a 6 layer Transformer model (~82 million parameters), a compressed version of the larger \roberta{} model.
\end{itemize}

\subsubsection*{Domain-specific models} 
We pre-train a range of Transformers on data from different domains and of different sizes, to gain insights on the effectiveness of pre-training on domain-specific data. One model (\stackobertflowshort{}) is entirely trained on domain-specific data, while the others are \enquote{off-the-shelf} models pre-trained on English, that are further pre-trained on some domain-specific data.

\begin{itemize}
  \item \bertreviews{}, a 12-layer (base) \bert{} model trained on \numprint{169097} (\SI{8.4}{\mega\byte}) unlabeled app reviews from the \arminer{} dataset.
  \item \bertcomments{}, a 12-layer (base) \bert{} model trained on \numprint{487693} (\SI{48}{\mega\byte}) comments extracted from well-known Java projects.
  \item \bertSOM{} and \bertSOMM{}, two 12-layer (base) \bert{} models trained on one (\SI{147}{\mega\byte}) and two millions
(\SI{304}{\mega\byte}) of Stack Overflow comments, respectively, taken from the \textit{Stack Exchange Data Dump}\footnote{\url{https://archive.org/details/stackexchange}}.
  \item \bertSOML{}, a 24-layer (large) \bert{} model trained on one million Stack Overflow comments (as above), used, however, only in the sentiment classification experiments. 
  \item StackOBERTflow, a 6-layer (small) \roberta{} model trained \textit{from scratch} on 26.2 million Stack Overflow comments (\SI{3.6}{\giga\byte}). The Stack Overflow corpus was tokenized using byte pair encoding (BPE) subwords and a large vocabulary size of \numprint{52000}.
\end{itemize}

\subsubsection*{Source code models}
For the code tasks, we use the following two models:
\begin{itemize}
    \item CodeBERTa\footnote{\url{https://huggingface.co/huggingface/CodeBERTa-small-v1}}, a small (6-layer)
\roberta{} model trained on the polyglot \textit{CodeSearchNet}~\citep{codesearchnet} source code corpus and released by Hugging Face.
The model supports multiple programming languages: Go, Java, Javascript, PHP, Python and Ruby. 
    \item CodeBERT~\citep{codebert}, a larger 12-layer model trained on the same corpus, but with a bimodal training regimen: the model takes as input pairs of natural language and code and primarily targets code-related tasks that also involve natural language (e.g., code search or summarization). Since none of our code experiments involves natural language, we use an empty string as natural language input, except for the code-comment coherence task, where, after stripping comment markers, comments are treated as natural language.
\end{itemize}

\subsubsection*{Scratch model}
To get a rough estimate of the effect of pre-training, we also train, for the sentiment classification task, a randomly initialized small \roberta{} model solely on the training set.

\subsection{Preprocessing}

Pre-trained models do not require extensive pre-processing, such as stemming or removing stop words. In fact, these may be harmful to performance, as the models were pre-trained on data that was \emph{not} pre-processed. In addition, large neural networks have enough parameter capacity to pick up on subtleties such as word order and negation.
Thus for most of the datasets, in line with the practice, we did  very little preprocessing. For the sentiment analysis and app review datasets, we used raw, unpreprocessed input. For the app review classification tasks we concatenated the review title and body and prepended the review's rating (a number in the range 1-5).
We applied heavier preprocessing for tasks with code comment input. Here, similarly to~\citep{satd-orig}, we removed newlines,
comment delimiters (such as \texttt{//}, \texttt{/*}, \texttt{*/}),
stripped HTML tags and removed all punctuation except periods and question marks as well as repeated whitespace characters.
Our goal was to reduce the length of the comments to fit in the Transformer's input window: each punctuation mark
is treated as an additional token, taking away a spot in the window.

Preprocessing was also necessary for some source code tasks. In the code-comment coherence task, we simply took the concatenation of the lead comment and the method body
as input. Moreover, we reformated all code files in the complexity prediction dataset using Google's Java code formatter\footnote{\url{https://github.com/google/google-java-format}}
such that, for instance, all of them use the same indentation width.
For the remaining dataset and tasks no preprocessing was done.

After this, we used each pre-trained model's tokenizer to properly segment the data in the subword units specific to this model, as each model may have a different vocabulary.

\subsection{Dataset Partitioning and Evaluation}
We tried to replicate the baseline models' training and evaluation methodology as closely as possible. Independent of the different evaluation strategies, we repeat all of our experiments at least three times with varying random seeds and average results to reduce noise.

\paragraph*{Sentiment classification} We trained our models on the predefined training set of the Senti4SD dataset \citep{senti4sd}, 30\% of which we use
for validation; the corresponding test set was used for testing. All the remaining sentiment analysis datasets were solely used as test sets (as was done in previous work).
Whenever a test set lacked a neutral sentiment class, as was the case for the JIRA issues dataset, we treated \textit{neutral} predictions from the model as negative.

\paragraph*{Informative app reviews} A predefined train-test split is also given for the informative app reviews detection task. Here, the test set is actually larger than the training 
set (\numprint{2000} and \numprint{1000} samples, respectively). We used 15\% of the training set for validation.

\paragraph*{App review classification} We used Monte Carlo cross-validation: we split the dataset in
10 random training and validation partitions with a ratio of 70:30. Reported results are averages over 10 runs. 

\paragraph*{SATD} We use cross-validation, with a 9 $\rightarrow$ 1 cross-project setting: we train on 9 out of the 10 total projects; the remaining project acts a test set. 

\paragraph*{Code-comment coherence} The model in \citep{corazza} was trained on 75\% of the dataset while the remaining 25\%
were used for testing. Because this train-test split was chosen randomly, an exact comparison is not possible.
To obtain more stable performance metrics, we re-evaluated this baseline model with three random train-splits and averaged the results. We train our own model in the same way, and use 10\% of the training data for validation.

\paragraph*{Linguistic code smells} As pointed out in the initial work, the leave-one-out cross validation strategy used for training the linguistic smell detection baselines is prohibitively expensive for a deep neural network.
We resorted to 15-fold cross validation, putting the baselines at a slight advantage: they were trained on over 99\% of the entire dataset, while our model uses only 93\%  of the data.

\paragraph*{Other datasets} Finally, the code runtime complexity prediction, comment classification and code readability prediction datasets were evaluated using $k$-fold cross-validation with $k$ equal to 5, 10 and 10 respectively.

\subsection{Fine-Tuning and Testing}

We tried several different hyper-parameters on the Senti4SD dataset, varying learning rate ($2\text{e-}5$, $4\text{e-}5$, $5\text{e-}5$, $5\text{e-}5$),
 batch size ($8$, $12$, $16$, $32$), and drop-out rate ($0.05$, $0.07$, $0.1$).
We found that general recommendations give good results in most cases (e.g., a learning rate of $5\text{e-}5$ for fine-tuning \bert{}). Interestingly, on Senti4SD a configuration
with a relatively small batch size of 12 worked best. On the remaining tasks batch size was selected so as to fill the available GPU memory (16-48 depending on GPU and task). This means that for medium-sized models, the batch size was halved with respect to small models, such as \codeberta{} or \stackobertflowshort{}.
We also reduced the Transformers input window size from 512 to 256 tokens on datasets where input sentences where so short that
the bulk of them fit this narrower window; a smaller window reduces memory consumption (and thus allows for a larger batch size) and speeds up training.
Exceptionally long samples that occurred occasionally were truncated to the used window size.
We stopped training after validation performance converged, which usually happened after 4-6 epochs.

\section{Results and Discussion}\label{sec:results}

 We first start by giving an overview of the performance on each task compared to the available baseline, before diving into the details of the impact on the performance of each of the techniques that we investigate. Of note, due to limitations in the datasets and the run-time needed for each experiment, we  were limited in the number of experiments we could run for each additional technique. For an overview of our results, refer to Table~\ref{tab:task-overview}. 

\begin{table}
  \footnotesize
  \begin{center}
  {\def\arraystretch{1}\tabcolsep=3pt\setcellgapes{8pt}
  \renewcommand{\arraystretch}{1.7}  
  \begin{threeparttable}[para]
  \begin{tabular}{|Cc|c|B|c|B|c|B|c|B|}
    \hline
    \makecell[c]{Sentiment Classification\\\cite{senti4sd, twodssenti, howfarcanwego}} & \hyperref[tab:senti-results]{\CIRCLE} & \improv & \noimprov & \lowimprov & && \improv & \\
    \makecell[c]{Informative App Review\\Detection~\cite{arminer}} & \hyperref[tab:ar-miner-results]{\CIRCLE} & \improv &&&&&& \\
    \makecell[c]{App Review\\Classification~\cite{review-classification, clap}} &  \hyperref[tab:review-classification-results]{\CIRCLE} & \improv & \noimprov & \improv && && \\
    \hline
    \makecell[c]{Self-Admitted Technical\\Debt Detection~\cite{satd-orig}} & \hyperref[tab:satd-results]{\CIRCLE} & \noimprov &&& \improv &&&\\
    \makecell[c]{Comment Classification\\\cite{comment-classification}} & \hyperref[tab:comment-classification-results]{\CIRCLE} & \noimprov &&&&&& \\
    \hline
    \makecell[c]{Code-Comment Coherence\\Prediction~\cite{corazza-ds}} & \hyperref[tab:coherence-results]{\LEFTcircle} &&&&& \improv && \\
    \makecell[c]{Linguistic Smell Detection~\cite{codesmell}} & \hyperref[tab:smell-detection-results]{\Circle} &&&&&&& \\
    \makecell[c]{Code Runtime Complexity\\Classification~\cite{complex}} & \hyperref[tab:complex-results]{\CIRCLE} &&&&&&& \\
    \makecell[c]{Code Readability\\Prediction~\cite{scalabrino2016}} & \hyperref[tab:readability-results]{\Circle} &&&&&&& \noimprov \\
    \hhline{-~|*{3}{>{\arrayrulecolor{Gray}}->{\arrayrulecolor{black}}|~|}>{\arrayrulecolor{Gray}}->{\arrayrulecolor{black}}|}
    \multicolumn{1}{|l}{Outcome} &&&&&&&& \\    
    \hhline{--*{3}{>{\arrayrulecolor{Gray}}->{\arrayrulecolor{black}}|~|}>{\arrayrulecolor{Gray}}->{\arrayrulecolor{black}}|}
    \multicolumn{2}{|l}{\cellcolor{Gray}\hyperref[subsec:pretrain]{Domain-specific pre-training}} &&&&&&& \\
    \hhline{---*{3}{~|>{\arrayrulecolor{Gray}}->{\arrayrulecolor{black}}|}}
    \multicolumn{3}{|l}{\hyperref[subsec:act-learn]{Active Learning}} &&&&&& \\
    \hhline{----*{2}{>{\arrayrulecolor{Gray}}->{\arrayrulecolor{black}}|~|}>{\arrayrulecolor{Gray}}->{\arrayrulecolor{black}}|}
    \multicolumn{4}{|l}{\cellcolor{Gray}\hyperref[subsec:backtrans]{Back-translation Augmentation}} &&&&& \\
    \hhline{-----*{2}{~|>{\arrayrulecolor{Gray}}->{\arrayrulecolor{black}}|}}    
    \multicolumn{5}{|l}{\hyperref[subsec:self-train]{Self-Training}} &&&& \\
    \hhline{------>{\arrayrulecolor{Gray}}->{\arrayrulecolor{black}}|~|>{\arrayrulecolor{Gray}}->{\arrayrulecolor{black}}|}
    \multicolumn{6}{|l}{\cellcolor{Gray}\hyperref[subsec:interm-train]{Intermediate-Task Fine-tuning}} &&& \\
    \hhline{-------~|>{\arrayrulecolor{Gray}}->{\arrayrulecolor{black}}|}
    \multicolumn{7}{|l}{\hyperref[subsec:soft-labels]{Label-Smoothing \& Soft-Labels}} && \\
    \hhline{-------->{\arrayrulecolor{Gray}}->{\arrayrulecolor{black}}|}
    \multicolumn{8}{|l}{\cellcolor{Gray}\hyperref[subsec:task-tokens]{Task-Specific Tokens}} & \\
    \hline

  \end{tabular}
  \begin{tablenotes}
  \item[] \makebox[\linewidth][s]{\makebox{\Circle = below baseline} \makebox{\LEFTcircle = close to baseline} \makebox{\CIRCLE = above baseline}}
  \item[] \makebox[\linewidth][s]{\makebox{\noimprov = no clear benefit}  \makebox{\lowimprov = little benefit} \makebox{\improv = likely benefit}}  
  \end{tablenotes}
  \end{threeparttable}
  }
  \end{center}
  \caption{Overview of tasks, experiments, and results in this work.} 
  \label{tab:task-overview}
\end{table}

\subsection{Comparison with Baselines}

\def\bertcommentscell{\thead{BERT-\\comments}}
\def\bertSOMcell{\thead{BERT-\\SO-1M}}
\def\bertbasecell{\thead{BERT\\(base)}}
\def\stackobertflowcell{\thead{Stack-\\OBERTflow}}

\paragraph*{Sentiment Analysis}
On the sentiment analysis datasets, Transformers are ahead of previous methods on most datasets (Table \ref{tab:senti-results}). In particular, this is
also true for the slightly out-of-domain datasets, such as the JIRA dataset, which the Transformers were not directly trained on, with one exception:
Transformers lag behind SentiStrength on the second Stack Overflow test set, but only in terms of F1 (by less than 1\%), not accuracy.

\paragraph*{App Review Analysis}
In both, app review classification (Table~\ref{tab:review-classification-results}) and informative app review detection (Table~\ref{tab:ar-miner-results})
Transformer models clearly outperform baselines. The BERT model that was further pre-trained on app reviews is in the lead, but all of the Transformers
manage to improve upon baselines. Finally, also on the CLAP dataset \stackobertflow{} was able to achieve a 5\% higher macro F1 score over
the previous random forest model proposed by~\citet{clap} (Table~\ref{tab:backtrans-results}).

\paragraph*{SATD}
The Transformer models are able to outperform the \textit{CNN} model by \citet{satd-cnn}, which, to the best of our knowledge,
represents the current state of the art (see Table \ref{tab:satd-results}). The dataset contains a considerable number of exact
duplicates and near duplicates (those arising after preprocessing): we report results with and without removal of such duplicates; we
do not know whether baseline have been trained with or without such duplicates.

\paragraph*{Comment Classification}
The Naive Bayes baseline lags behind all three Transformer models (\stackobertflow{}, standard BERT, and a domain-specific pre-trained BERT)
by a margin of 4\% (Table~\ref{tab:comment-classification-results}). A BERT model further pre-trained on task-specific data (i.e., Java comments)
performed slightly worse than standard BERT.

\paragraph*{Code-Comment Coherence}
The SVM baseline by \citet{corazza} performs better than the CodeBERTa Transformer, even
when employing intermediate-task training (+1\% accuracy, Table \ref{tab:coherence-results}).
In a later work, \citet{corazza-we} experiment with word embeddings: the resulting baseline
is weaker than their first and outperformed by the Transformer.

\paragraph*{Linguistic Smell Detection}

We compare with the baselines established by \citet{codesmell} in Table~\ref{tab:smell-detection-results}. CodeBERTa is able to outperform the manually tuned SVM and 
the also CNN but clearly remains behind the SMO (sequential minimal optimization) model 
that was automatically tuned using Bayesian optimization (through Auto-Weka~\citep{auto-weka}).

\paragraph*{Runtime Complexity Classification}
The complexity classification task was the only code task where the Transformer exceeded all baselines (Table~\ref{tab:complex-results}),
including the Random Forest classifier and the SVM trained on AST embeddings.

\paragraph*{Code Readability Prediction}
The logistic regression baseline trained on manually engineered features by \citet{scalabrino2018} is out of reach
for the Transformer: the accuracy achieved by the baseline is over 10\% higher (Table~\ref{tab:readability-results}).
From all the selected tasks, the readability prediction task was the hardest for the Transformer.

\begin{table}[!tbph]
  \begin{center}
  {\def\arraystretch{1}\tabcolsep=3pt\setcellgapes{3pt}  
  \begin{tabular}{lccccc}
   \toprule
   \textbf{Model} & \textbf{\thead{Face-\\book}} & \textbf{\thead{Tap\\Fish}} & \textbf{\thead{Temple\\Run2}} & \textbf{\thead{Swift-\\Key}} & \textbf{Avg.} \\
   \midrule
   \input{ar_miner/results/table}
   \bottomrule
  \end{tabular}
  }  
  \end{center}
  \caption{Results (macro F1) for four apps in the \arminer{} dataset. Our numbers are averages over five runs with different seeds.}
  \label{tab:ar-miner-results}
\end{table}  

\conclusion{Pre-trained transformers were able to outperform baselines on domains closer to natural language; for source code, results were mixed.}

\begin{table}[!tbph]
  \begin{center}
  \begin{threeparttable}
  \begin{tabular}{lccc}
   \toprule
   \textbf{Model} & \textbf{F1} &  \textbf{Prec.} & \textbf{Rec.} \\
   \midrule
   \input{smell_detection/results/table}
   \bottomrule
   \end{tabular}
   \begin{tablenotes}
    \item[1] 15-fold cross validation
    \item[2] leave-one-out cross validation
   \end{tablenotes}
   \end{threeparttable}   
  \end{center}
  \caption{Macro F1, precision and recall for the linguistic smell detection task. Our numbers are averages over three runs with different seeds.}
  \label{tab:smell-detection-results}
\end{table}  

\begin{table}[!tbph]
  \begin{center}
  \begin{tabular}{lr}
   \toprule
   \textbf{Model} & \textbf{Acc.} \\
   \midrule
   \input{corcod/results/table}
   \bottomrule
   \end{tabular}
  \end{center}
  \caption{Results for the code runtime complexity prediction task. Our numbers are means over five runs with different seeds.}
  \label{tab:complex-results}
\end{table}

\begin{table*}[!tbhp]
  \footnotesize
  \begin{center}
    {\def\arraystretch{1.0}\tabcolsep=1pt%
    \begin{tabular}{@{\extracolsep{4pt}}lcccccccccc@{}}
      \toprule
      {} 
      & \multicolumn{4}{c}{without duplicates}
      & \multicolumn{4}{c}{with duplicates}  
      &  &  \\
     \cline{2-5}
     \cline{6-9}

      {} & \bertSOMcell{} & \bertcommentscell{} & \bertbasecell{} & \stackobertflowcell{}
      & \bertSOMcell{} & \bertcommentscell & \bertbasecell{} & \stackobertflowcell{}
      & CNN & NLP \\

      \midrule
      \input{satd/results/table_f1}
      \bottomrule
    \end{tabular}
    }
  \end{center}
  \caption{Macro F1 scores for the SATD detection task. As far as our results are concerned, numbers are means over five runs, each with different seed.}
  \label{tab:satd-results}
\end{table*}

\begin{table}[!htbp]
  \begin{center}
  \begin{tabular}{lccc}
   \toprule
   \textbf{Model} & \textbf{F1} & \textbf{Prec.} & \textbf{Rec.} \\
   \midrule
   \input{comment_classification/results/table}
   \bottomrule
   \end{tabular}
  \end{center}
  \caption{Results for the comment classification task. Task-specific pre-training failed to improve performance. As far as our models are concerned, results are averages over three runs with different seeds.}
  \label{tab:comment-classification-results}
\end{table}  

\begin{table}[!htbp]
  \footnotesize
  \begin{center}
    {\tabcolsep=2pt%
    \begin{tabular}{@{\extracolsep{2pt}}llcc@{}}
      \toprule
       \textbf{Dataset} &  \textbf{Model} & \textbf{Acc.} &   \textbf{F1} \\
       \midrule
       \input{sentiment/results/condensed}       \bottomrule
    \end{tabular} 
    }
  \end{center}
  \caption{Accuracy, macro F1 and per-class precision and recall for different models and datasets. Values reported are means over five runs, each with different seed (only our models). All models were trained on the \dssentiforsd{} \stackoverflow{} dataset.} 
  \label{tab:senti-results}
\end{table}

\subsection{Pre-Training}

Our results suggest that, in particular for natural language tasks, the most promising approach seems to be to further pre-train models already pre-trained on general English.
When available, in-domain data should be used for pre-training, but even close-to-domain data can yield good improvements. For instance, pre-training \bert{} on Stack Overflow
comments helped to improve accuracy also on the app review dataset (Table \ref{tab:senti-results}). Further pre-training an already pre-trained model should also be preferred over pre-training from scratch: The further pre-trained models outperformed our model pre-trained from scratch for most tasks and metrics even though pre-training from scratch required considerably more training time and (unlabeled) training data. This comparison comes with a grain of salt: our further pre-trained models have twice as many layers as our pre-trained-from-scratch model, which, in turn, has a much larger vocabulary (\numprint{52000} vs \numprint{30522}). While not having a larger model pre-trained from scratch is a limitation of this work, it also highlights how expensive it is.

What speaks for our small model, and for small models in general, is of course their size: with only half the layers, training and evaluation is roughly twice as fast,
the memory footprint is much smaller and, depending on the task, the performance hit may be acceptable.

Our experiments also indicate that further pre-training is effective even with relatively small amounts data.
In the sentiment classification task as little as \SI{150}{\mega\byte} (1 million samples) of pre-training data seems to be sufficient and able to \enquote{saturate} the model. Doubling the amount of pre-training data resulted in virtually negligible improvements (see \bertSOM{} versus \bertSOMM{} in Table \ref{tab:senti-results}).
Similarly, for our large model (\bertSOML{}) improvements are marginal: on the Senti4SD test set, i.e., the test set that \enquote{matches} the training set, it
outperforms the base-sized models by only 0.6\%, while on the other test tests it lags behind them.

Interestingly, our \bertcomments{} model, a general English model further pre-trained on Java comments, performs slightly worse than the
same model without this task-specific pre-training (i.e., a standard \bert{}) on both datasets it was applied to (Tables \ref{tab:comment-classification-results} and \ref{tab:satd-results}).
As to why this is the case we can only \textit{speculate}: A possible explanation is that the comments in our pre-training dataset are very repetitive and have
low linguistic diversity (e.g., Java docstrings). Thus, the model might have unlearned some of its general language capabilities during task-specific pre-training. 

\begin{figure}[!htbp]
    \begin{center}
        \scalebox{0.7}{\input{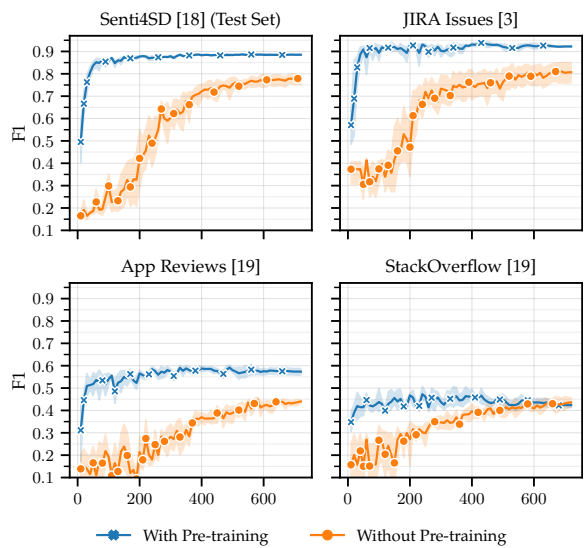}}
    \end{center}
    \caption{F1 score on different sentiment classification datasets with and without pre-training.
             Number of optimization steps is shown on the x-axis; error bands are 95\% confidence intervals.}
    \label{fig:senti-no-pretrain}
\end{figure}

Figure \ref{fig:senti-no-pretrain} demonstrates that pre-training is essential: a randomly initialized model not only converges much slower, it also has higher variance
and typically reaches much lower peak performance. In sum, our experiments show that there is very little reason not to use an already pre-trained, general natural language model as the basis for further domain-specific pre-training and should in most cases be preferred over pre-training from scratch, which, in relation
to training time, hardly seems worth the effort.

\conclusion{Pre-training is essential. Further pre-training a generic model on domain-specific data is often beneficial, and is much more effective than pre-training from scratch.}

\subsection{Soft Labels}
\label{subsec:soft-labels}

Since ~\citet{senti4sd} released multi-rater labels (three per sample) along with the majority label, we conducted a soft label experiment. %
For instance, if one voter assigned the \textit{positive} label to a sentence, while two raters assigned \textit{neutral}, majority voting would label it as \textit{neutral}.
Instead, the soft label captures all three rater labels, assigning the distribution: (positive: $0.\overline{33}$, negative: $0$, neutral: $0.\overline{66}$)
to the sentence.

\begin{table}[!htbp]
  \footnotesize
  \begin{center}
    {\tabcolsep=2pt
    \begin{tabular}{llr@{${}\pm{}$}rr}
      \toprule
      \multirow{2}{*}{\textbf{Dataset}} &  \multirow{2}{*}{\textbf{Label Type}} &  \multicolumn{3}{c}{\textbf{F1}}\\
      \cline{3-5}
                                        &                                       &  \multicolumn{2}{c}{$\mu \pm \sigma$} & $max$ \\
      \midrule                                        
      \input{sentiment/soft_label/condensedtable}
      \bottomrule
    \end{tabular}
    }
  \end{center}
  \caption{Macro F1 for different label types and datasets: mean, maximum and standard deviation over five runs, each with different seed.  All models were trained on the \dssentiforsd{} \stackoverflow{} dataset.}
  \label{tab:soft-labels}
\end{table}

We train a subset of the models with hard-labels and soft-labels on Senti4SD, and evaluate on all sentiment analysis datasets (see Table \ref{tab:soft-labels}). The results look
rather promising: on the Senti4SD
test set, training with all three rater labels resulted in an
increase of 0.5\%. On the JIRA test set, soft labels
yielded an improvement of 1\%. On the other hand, performance dropped on the second Stack Overflow test set. 
While these improvements are not certain, and might seem modest,
they come almost for free. Whenever multi-rater labels are available, we
recommend to tentatively use them in this way
and encourage creators of datasets to also release labels of individual raters.

\conclusion{When available, individual rater labels may improve performance at very low cost.}

\subsection{Back-Translation Augmentation}
\label{subsec:backtrans}

We performed back-translation experiments on the sentiment and app review classification tasks by
translating the entire datasets into French, German and Russian using Google Translate and from these
languages back into English (see Figure \ref{fig:backtans-example} for an example). 
For the CLAP and Senti4SD datasets we do training-time and test-time augmentation, both, separately and combined, using the StackOBERTflow model (Table \ref{tab:backtrans-results}).
On the other app review dataset \citep{review-classification} we do training-time augmentation alone, and combine it with test-time augmentation: here the the experiment is carried out on several different models (Table \ref{tab:review-classification-results}).

\begin{table}[!htbp]
  \begin{center}
    {\tabcolsep=2pt\footnotesize
    \begin{threeparttable}
    \begin{tabular}{@{\extracolsep{4pt}}lccccc@{}}
      \toprule
      {} & \textbf{\thead{Bug\\reports}} & \textbf{\thead{Feature\\request}} & \textbf{\thead{Ratings}} & \textbf{\thead{User\\experience}} & \textbf{\thead{Avg.}} \\
      \midrule
      \input{review_classification/results/table}
      \bottomrule
    \end{tabular}
    \begin{tablenotes}
      \item[1] multiclass, bag of words + metadata
    \end{tablenotes}
    \end{threeparttable}
    }
  \end{center}
  \caption{F1 scores for app review classification, reported with training-time back-translation augmentation (+BT), with training-time \emph{and} test-time back-translation augmentation (+BTT) and without any augmentation. Results are averages over three runs with different seeds.}
  \label{tab:review-classification-results}  
\end{table}

\begin{table}[!htbp]
  \begin{center}
    {\tabcolsep=4pt
    \begin{tabular}{llcccc}
      \toprule
      \textbf{Dataset} & \textbf{Augmentation} &  \textbf{Acc.} & \textbf{F1} & \textbf{Prec.} & \textbf{Rec.} \\
      \midrule
      \input{backtrans/table_clap}
      \midrule
      \input{backtrans/table_senti4sd}      
      \bottomrule
    \end{tabular}
    }
  \end{center}
  \caption{Macro F1, precision and recall for back-translation augmentation at training and test time on the Senti4SD (sentiment classification) and CLAP (app review classification) datasets using the StackOBERTflow model. Results are averages over three runs with different seeds.}
  \label{tab:backtrans-results}  
\end{table}

Back-translation augmentation led to
a clear increase in F1 and accuracy on the CLAP app review dataset, in particular when training and testing time
augmentation were combined (+1.1\% accuracy, Table \ref{tab:backtrans-results}). On Senti4SD, data augmentation
yields modest improvements (+0.3\%); in fact, augmenting at test time only caused a
slight drop in performance. Table~\ref{tab:review-classification-results} suggests that the effect of back-translation augmentation depends on the model and pre-training choice. With train-time augmentation only, we see a modest increase of 0.6\% in F1 for our small StackOBERTflow model and 1.1\% on a general \bert{} model,
while \bertreviews{} shows better performance without augmentation. The latter does however benefit from combined augmentation (+0.4\%). However, the question of whether in general
task-specific pre-training diminishes the effects of data augmentation cannot be answered given this limited data and would require further experiments.

\conclusion{When possible, back translation  yields improvements, particularly if used at both training and test time.} 

\subsection{Active Learning}

\label{subsec:act-learn}

\def\Dtest{\mathcal{D}_{test}}
\def\Dpool{\mathcal{D}_{pool}}
\def\Dseed{\mathcal{D}_{seed}}
\def\Dtrain{\mathcal{D}_{train}}
\def\Dtop{\mathcal{D}_{top}}

\begin{table}[!htbp]
  \small
  \begin{center}
    {\def\arraystretch{1.6}\tabcolsep=2pt  
    \begin{tabular}{r@{=}l}
       $C_{LC}(x)$ & $\frac{n}{n-1}(1 - P_\theta(y^{*}_{1}|x))$ \\
       $C_{MC}(x)$ & $1 - (P_\theta(y^{*}_{1}|x) - P_\theta(y^{*}_{2}|x))$ \\
       $C_{RC}(x)$ & $\frac{P_\theta(y^{*}_{1}|x)}{P_\theta(y^{*}_{2}|x)}$ \\
       $C_{E}(x)$ & $-\frac{1}{log_2(n)}\sum_{i=1}^n{P_\theta(y_{i}|x)log_2(P_\theta(y_{i}|x))}$ \\
       $C_{rand}(x)$ & $rand([0, 1])$ \\
    \end{tabular}
    }
  \end{center}
  \caption{Acquisition functions used in our active learning experiment, adapted from \citep{humanintheloop}: least confidence ($C_{LC}$), margin of confidence ($C_{MC}$), ratio of confidence ($C_{RC}$), entropy ($C_{E}$) and random confidence ($C_{rand}$). $y^{*}_{1}$ and $y^{*}_{2}$ are the classes with highest and second highest probabilities, respectively; $n$ is the number of classes. All functions have range $[0, 1]$.}
  \label{tab:acqfuncs}
\end{table}

We try active learning on the Senti4SD sentiment analysis dataset and an app review dataset. In both
cases we compare several acquisition functions (Table \ref{tab:acqfuncs}).
We carry out the experiment as follows: initially we split the \textit{training} set into the seed set $\Dseed$ containing 5\% of all samples and
the pool set $\Dpool$, containing the remaining samples. We let $\Dtrain := \Dseed$ and train the model. Then we evaluate $\Dpool$ as well as the test set on this model.
Next, for each $x \in \Dpool$ we calculate a confidence score by applying the acquisition function (from Table \ref{tab:acqfuncs}).
After that, we let $\Dtop$ be the $k=180$ samples with the highest confidence score. We remove these samples from $\Dpool$ %
and add them to $\Dtrain$ %
This procedure is repeated until $\Dpool$ is empty.

\begin{figure}[!htbp]
    \centering
    \scalebox{0.65}{\input{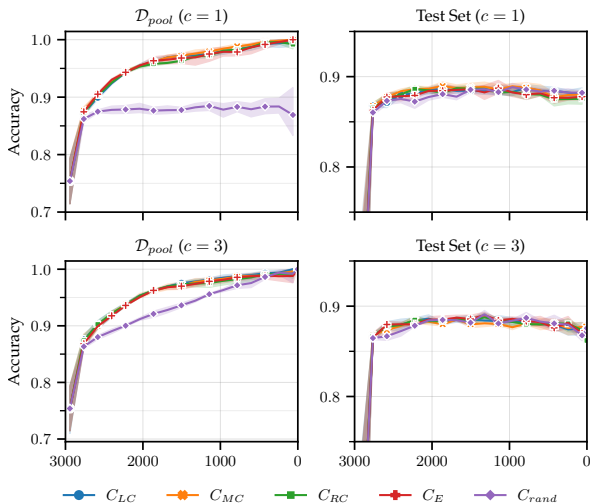}}
    \caption{Accuracy of $\Dpool$ and the test set at each iteration of the active learning process for different acquisition functions with ($c=3$) and without ($c=1$) diversity sampling for the sentiment classification task (Senti4SD). Error bands are 95\% confidence intervals. Same plot for the review classification task can be found in the appendix (Figure \ref{fig:act-learn-appendix})}
    \label{fig:act-learn}
\end{figure}

We also combine confidence sampling with diversity sampling to avoid introducing similar samples into the pool. Instead
of $k$ samples, we select $c \cdot k$ samples from the pool, where $c$ determines the cluster size. We use the \textit{k}-means algorithm to cluster the $c \cdot k$ samples 
into $k$ clusters, each of size $c$. We then select a single sample from each cluster, for a total of $k$ samples. Then, we proceed as above. 

Figure \ref{fig:act-learn} shows evaluation results at each iteration step with a cluster size of $c=3$ for the sentiment classification task. %
The outcome of our active learning experiments remained behind expectations: in both tasks,
neither confidence sampling alone nor confidence and diversity sampling combined showed an appreciable advantage over the random baseline.  The choice of acquisition function  did not seem crucial, but a more systematic study would be needed to draw more solid conclusions. On the other hand, the plot of pool accuracy (top left) indicates that the active learning process worked as expected: a random acquisition function without diversity sampling had constant performance. %

\conclusion{Active learning simulations did not improve results.}

\subsection{Self-Training}
\label{subsec:self-train}
We investigate the use of self-training  for the SATD detection task.
We extract \numprint{350000} comments from various popular Java libraries and frameworks.
Then we train a classifier model on the entire \textit{original} dataset, which we use to classify the \numprint{350000} comments as
either \textit{technical debt} or \textit{not technical debt}.
We only keep the \numprint{7904} positive comments, i.e., those classified as \textit{technical debt}, and discard all other samples to avoid increasing the class imbalance already extant in the original dataset.
For each positive sample we calculate a confidence score using $C_{LC}$ (Table \ref{tab:acqfuncs}).
We take the top 5\%, and 80\% most confidently classified comments, equal to \numprint{6092} and \numprint{7880} additional comments, respectively, and add them to the original training set.
Finally, the model is trained and evaluated on this extended training set. 

\begin{figure}[!htbp]
    \begin{center}
        \scalebox{0.55}{\input{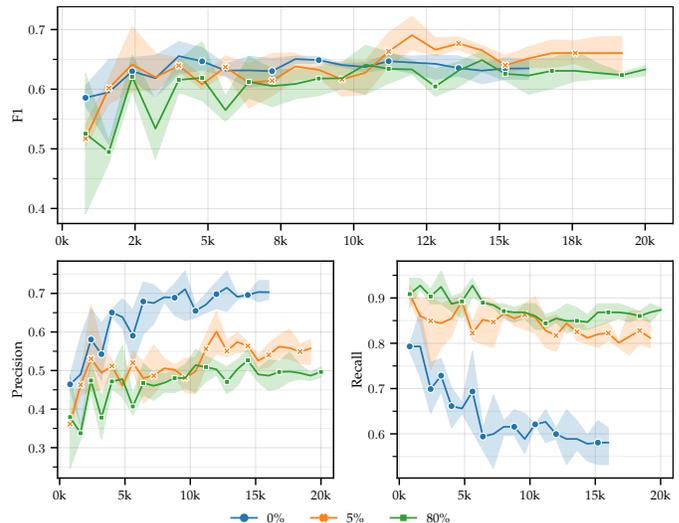}}
    \end{center}
    \caption{Precision, recall and F1 under different self-training settings for the \textit{Apache Ant} project. Error bands are 95\% confidence intervals.}
    \label{fig:self-train}
\end{figure}

Figure \ref{fig:self-train} shows that self-training increases recall (and possibly F1) but causes precision to drop.
This is, of course, not surprising: the added samples increase dataset variance which likely explains an increased recall. Similarly, the precision drop
can be explained by the lower quality self-training labels. %
Thus, one can tune the precision-recall trade-off according to task-specific needs, such as when a recall is more important than precision, or the model's precision is high enough to be partly sacrificed for better recall. Figure \ref{fig:self-train} shows different self-training settings for \textit{Apache Ant}: F1 score went up 4\% (precision: -1\%, recall +10\%), when using a 5\% confidence threshold. The change in F1 strongly depends on the
confidence threshold and varies across projects:  \textit{EMF} sees an 8\% F1 drop (precision: $-28$\%, recall: $+15$\%); other projects range from $-1$\% to $+4$\%. 

\conclusion{Self-training increases recall at the expense of precision, but the confidence threshold should be tuned carefully.}

\subsection{Intermediate-Task Training}
\label{subsec:interm-train}

We evaluate Intermediate-Task Training (ITT) on the code-comment coherence task, as it is the only setting for which we could define such an intermediate task. We use \numprint{38000} Java methods along with their lead comments from the \textit{CodeSearchNet} \citep{codesearchnet} dataset. We assign half of the methods to their actual lead comments (assumed to be coherent) and shuffle the other half randomly (thus assumed to be incoherent).
The model is then fine-tuned on the intermediate task of detecting whether a method was paired with its true lead comment or a random one. Finally, we fine-tune on the code-comment coherence dataset as usual.

\begin{table}[!htbp]
  \begin{center}
    {\tabcolsep=16pt
    \begin{threeparttable}
    \begin{tabular}{lccc}
      \toprule
      \textbf{Model}  &  \textbf{Acc.} & \textbf{AUC} \\
      \midrule
      \input{coherence/results/table}
      \bottomrule
    \end{tabular}
    \begin{tablenotes}
      \item[1] Single seed; value reported in \cite{corazza}
      \item[2] Average over 5 seeds; reproduction
    \end{tablenotes}
    \end{threeparttable}
    }
  \end{center}
  \caption{Results on the comment-code coherence dataset. \codeberta{} was evaluated with and without intermediate-task training (ITT). Our numbers are averages over five runs with different seeds.}
  \label{tab:coherence-results}  
\end{table}

ITT improved the performance of the Transformer model on the code-comment coherence task (Table \ref{tab:coherence-results}): we observed a modest rise in AUC of 0.8\%, and a slightly higher increase in accuracy~(+1.1\%). If an appropriate intermediate task for the task at hand can be found, ITT can be done with relatively little effort: in our case, the training procedure for the intermediate-task was mostly identical to the one for the target task; the bulk of the work consisted in generating the intermediate-task dataset (e.g. selecting, shuffling and pre-processing the data from \textit{CodeSearchNet}).

\conclusion{When applicable, ITT may improve performance; however finding a suitable intermediate task may be difficult.}

\section{Discussion}\label{sec:limit}

\subsection{Summary and implications of the results}

\paragraph*{Types of datasets} Overall, we see that Transformers work very well for natural language datasets, but that performance on source code is ``hit or miss''.  This comes with caveats: the models used so far are multilingual, which might reduce performance. They are also trained with less computational resources, and on an order of magnitude less data: While CodeSearchNet is around 1.7 GB, the training data for BERT is around 16GB, while it is 160GB for RoBERTa (CodeSearchNet). Models where also small (6 layers), or had a dual input (text and code). We would expect a Java-specific model trained on a similar size corpus as BERT to perform better. Moreover, source code is quite different from natural language: code snippets are often larger than sentences, and much more structured. This might pose limits on what an unstructured model might achieve. Recent adaptations of the Transformer architecture (e.g. ~\citep{trees-to-transformers, pos-enc-transformers}) allow the model to better make use of the tree-like structure of code. Investigating such code-specific architectures in connection with small datasets remains an issue for future research.

\paragraph*{Domain-specific pre-training} proved effective in natural language settings, improving performance at a moderate cost in terms of computation and data. The only case where it did not work well was for code comments. While we are not sure why, one reason could be that code comments are too far way from regular English (needing a specific model instead), or that careful curation of the data set (avoiding too many duplicates) is needed. Both cases could lead to catastrophic forgetting~\citep{cata-forgettig} of the initial pre-training. Leveraging the resources that were used to train BERT and fine-tuning it further proved much more effective than training a model from scratch.
We have not evaluated domain-specific training from English to  source code, as we hypothesized that the two domains are very different---the tokenization alone might differ significantly \cite{karampatsis2020big}. This intuition is supported by the literature, which reports an example where an English BERT was applied to source code, with underwhelming results \citep{bert-arch-tactics}. %

\paragraph*{Back-translation} While data augmentation is effective for natural language, it is not immediately applicable to source code. Source code can not be “back translated” easily. Specific data augmentations for code should be investigated, but may not be trivial (e.g., renaming identifiers could be investigated, but what should be done with API methods?).

\paragraph*{Intermediate Task Training} Another alternative is to define suitable intermediate training tasks. We have found initial evidence of this, and a recent paper adds further evidence, in the context of traceability~\citep{traceability-transformed}. However, it used a very similar task and dataset. Thus, the challenge here is not whether intermediate task training helps, but rather \emph{whether a suitable task exists} for a given problem.

\paragraph*{Soft Labels} Soft labels that reflect the uncertainty of raters (and thus the difficulty of the samples) can be useful as well, and at a minimal cost. However these are not common, as of now. We call on dataset builders to release them alongside the majority label, as was done by Calefato \cite{senti4sd}.

\subsection{Limitations of this work}

\paragraph*{Limited Number of experiments} While we try to report results as extensively as possible to increase their generalisability, we are limited for two main reasons:
\begin{inparaenum}[1)]
\item we have limited computational resources, and
\item some techniques are specific to some settings.
\end{inparaenum}

\paragraph*{Limited resources} Deep learning is famously resource intensive. While fine-tuning is less resource intensive than training models from scratch, it still requires significant time on one or more dedicated GPUs, particularly for larger models. A single run is measured in hours. This limits the number of experiments, particularly as we repeat experiments several times with different random seeds. %

\paragraph*{Results in specific settings} While resources are limited, we still wanted to try each technique on at least two datasets. However, some techniques were applied to a single dataset. For soft labels, we needed multiple ratings: only a single sentiment analysis dataset had the required three ratings per sample. We could only define a reasonable intermediate task for code-comment readability prediction. We considered using self-training for comment classification, but did not, due to the large number of imbalanced classes.

\paragraph*{Hyper-parameters} Limited resources also impact the extent to which we perform hyper-parameter optimisation, as thorough parameter searches (whether by grid, random or bayesian methods) would be prohibitively expensive. A second limit is that some hyper-parameters are fixed by the usage of a pre-trained model (e.g. number of layers, number of attention heads, embedding size, vocabulary size). A silver lining is that, given the interest in pre-trained models, general recommendations for hyper-parameters exist and are broadly applicable. %
Thus, we started with these recommendations, and investigated some variations of the hyper-parameters on the Senti4SD dataset, confirming that the recommendations worked well. We then applied those hyper-parameters on other experiments, varying only the most important ones in some cases (learning rate, batch size). Cross-validation also makes evaluation and hyper-parameter tuning more complex and resource intensive.  Since we limited hyper-parameter tuning, we are not at risk of overfitting to the test fold when doing cross validation. An alternative would be to use doubly nested cross validation, but this further increases the resource needed. We note that dedicated test sets ease this considerably. %

\paragraph*{Comparisons with previous work} We do our best to provide a fair comparison with previous work, while avoiding methodological issue (e.g. averaging seeds). We do not always exactly know how previous work was evaluated evaluation (e.g., hyper-parameter selection strategy, whether simple or nested cross-validation was used, or whether some data points were excluded) as code is not always released. In some other cases, other factor presents us to make an exact comparison (e.g. use of leave-one-out cross validation is not practical for our setting). To alleviate this in the future, we release our source code (see Appendix~\ref{appendix:implementation}).

\paragraph*{Active Learning} While we could not see an advantage to active learning, this is not in line with previous work by \citet{review-classification-active-learn} and \citet{tu2020better}.Of note, our results are obtained through simulation based on an existing labelled dataset. While this is a practice often used to evaluate active learning methods, a realistic application of active learning on a larger set of unlabelled data would lead to a different training set, which may be substantially more varied, and thus more effective. But it is also possible that the impact of active learning is less visible when pre-training is used. %

\paragraph*{Random Seeds} 

\citet{bertseeds} found that the choice of the random seed can have a substantial impact on performance, especially for small datasets.
We ran most of our experiments five times and all of them at least three times, with different seeds.
While this surely mitigates the problem, it might not fully clear it up .%

\paragraph*{Implementation Bugs}

Our implementations are based on Hugging Face's \texttt{transformers} Python package \cite{wolf2020huggingfaces}, a high quality implementation of common Transformer models. However, despite careful reviews we cannot fully preclude errors in our own code and adaptations.

\section{Conclusions}\label{sec:conc}

Software Engineering datasets are often small, by necessity. In this work, we trained various Transformer models on 13 small and medium-sized dataset selected from the recent Software Engineering literature.
We not only compared Transformers of different size and different pre-training regimes but also applied several
machine learning techniques that promised a possible benefit for small datasets. These techniques were
data augmentation, self-training, intermediate-task training, active learning and soft labels. %

Overall, we found that on natural language tasks, Transformers usually outperform existing baselines. On source code tasks, however, results were mixed. Significant work lies ahead to define effective pre-trained source code models either by training larger models on more data, or by incorporating more structural information during training.

In general, we advise \emph{against} pre-training a new model from scratch as it is extremely resource intensive, for mixed results.  Instead, an already pre-trained model can be \emph{further pre-trained} on task-specific data. If such task-specific data is unavailable, training on close-to-domain data is worth a try. We provide several such pre-trained models in Appendix~\ref{appendix:implementation}.

Several additional techniques were useful at a relatively low cost. We particularly recommend the use of soft labels derived from multi-rater labels if available, and call on dataset authors to release these multi-rater labels.
Back-translation is similarly useful, if more expensive. It is unfortunately not easily applicable to source code. 

Other techniques were less applicable. We find that self-training is advisable only in cases where the user wants to boost recall and is willing to sacrifice precision. If circumstances allow it, intermediate task training seems promising, but it seems rarely applicable, and has a much higher cost. Finally, our active learning experiments were inconclusive; a wider study on a larger set of dataset might be required to draw a clearer picture. 

While these general guidelines are useful on their own, their applicability is limited. To this extent, we release all the scripts and pretrained models that were built as part of this work, so that the community can easily fine-tune the models on their own Software Engineering datasets, and apply additional techniques as they see fit (see Appendix~\ref{appendix:implementation}).

\printbibliography

\FloatBarrier
\begin{IEEEbiography}[{\includegraphics[width=1in,height=1.25in,clip,keepaspectratio]{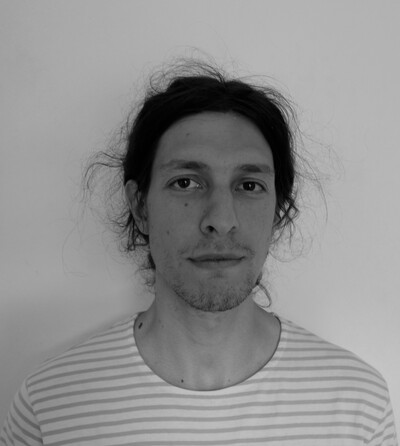}}]{Julian Aron Prenner}
received his M.Sc. in Computer Science from the Free University of Bozen-Bolzano, where he is also currently pursuing his Ph.D. His research interests include program repair, automatic test case generation and applications of machine learning in Software Engineering.
\end{IEEEbiography}

\begin{IEEEbiography}[{\includegraphics[width=1in,height=1.25in,clip,keepaspectratio]{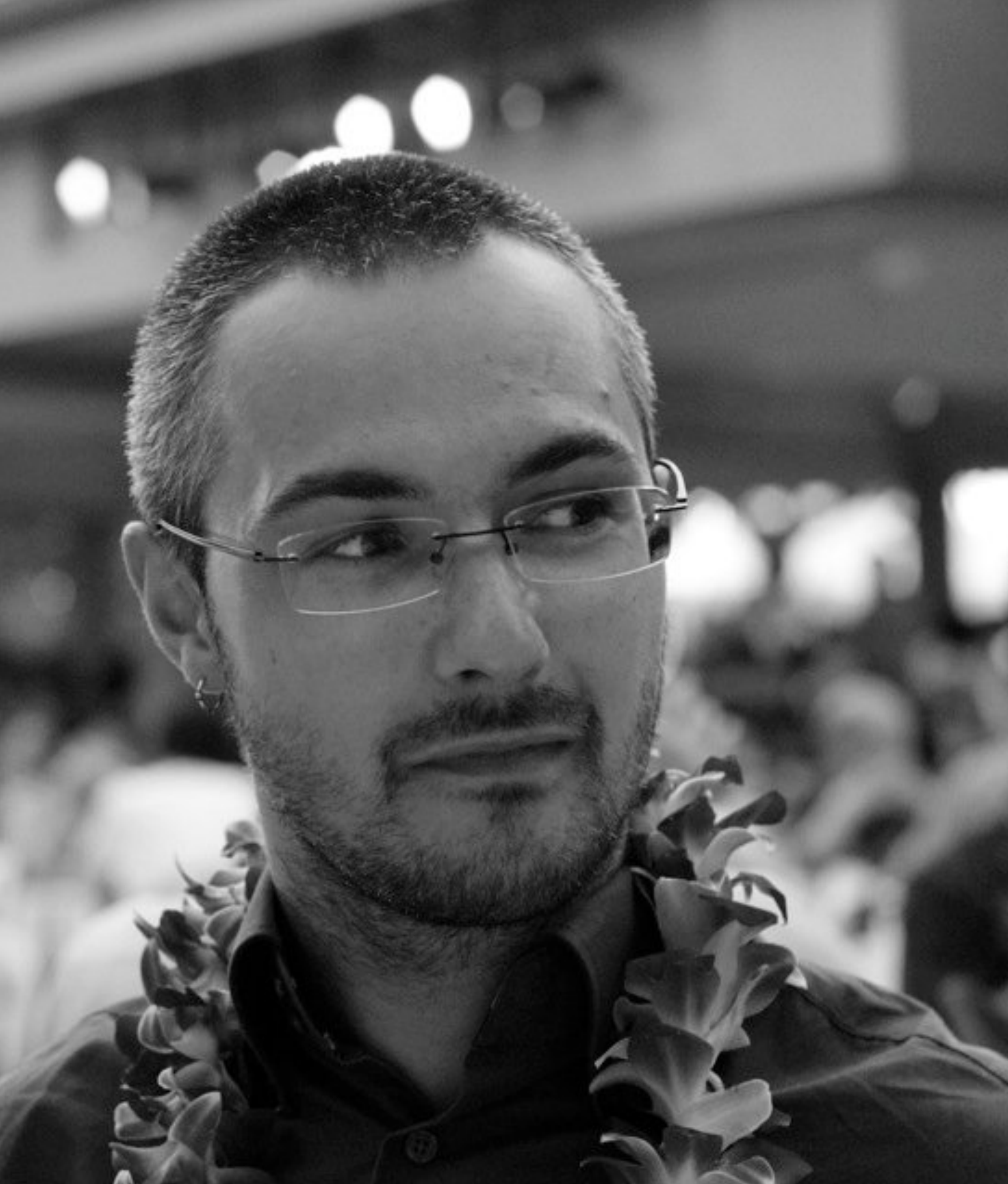}}]{Romain Robbes}
is an Associate Professor at the Free University of Bozen-Bolzano, in the SwSE research group. Before that, he was an Assistant, then Associate Professor at the University of Chile. He earned his PhD in 2008 from the University of Lugano, Switzerland and his Master’s degree from the University of Caen, France. His research interests include Empirical Software Engineering, Software Evolution, Mining Software Repositories, and Machine Learning for Software Engineering. 
\end{IEEEbiography}

\newpage
\appendices

\begin{figure}[!htbp]
  \centering
  \scalebox{0.65}{\input{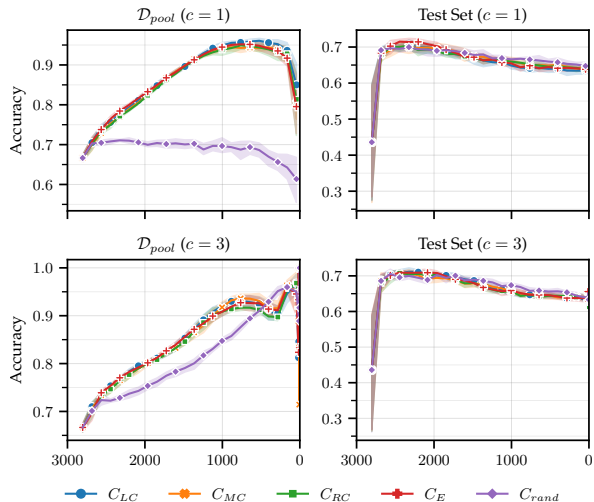}}
  \caption{Accuracy of $\Dpool$ and the test set at each iteration of the active learning process for different acquisition functions with ($c=3$) and without ($c=1$) diversity sampling for the review classification task. Error bands are 95\% confidence intervals.}
    \label{fig:act-learn-appendix}
\end{figure}

\section{Additional information on the datasets}
\label{appendix:datasets}

\paragraph*{Data Quality}

On the SATD dataset we noticed that false positives often contain keywords such as \texttt{FIXME}, \texttt{TODO}, or \texttt{HACK};
while this might raise questions about  quality, we found that only 1.2\% of the negative instances contain such keywords,
compared to 6\% among positives.
\AtNextCitekey{\defcounter{maxnames}{1}\defcounter{minnames}{1}}
Similarly, in the app review dataset by \citet{review-classification}
we found that 9\% of the samples appear twice, with different labels. We left these duplicates 
in the dataset as we did not know how this issue was handled by previous work, nor
which duplicates to remove and which to keep.

\paragraph*{Additional examples and Statistics}

Table~\ref{tab:ds-examples} shows a representative example of each dataset, alongside with its class. Table~\ref{tab:satd-ds} shows detailed statistics (per project) on the Self-Admited Technical Debt Dataset.

\input{samplestable}

\begin{table}
  \footnotesize
  \begin{center}
    \input{satd/dataset/table}
  \end{center}
  \caption{Projects in the SATD dataset along with the number of samples classified as self-admitted technical debt.} 
  \label{tab:satd-ds}
\end{table}

\section{Additional results}
\label{appendix:results}

\subsection{Active Learning}
Figure \ref{fig:act-learn-appendix} shows the results of active learning on the app review classification dataset.
On this dataset, all pool curves show a noticeable drop at the end. A possible explanation for this might be data quality: we observed that the \AtNextCitekey{\defcounter{maxnames}{1}\defcounter{minnames}{1}} app review dataset by \citet{review-classification} contains a number of duplicates with conflicting labels. Having learned one of the duplicate samples, all its copies will be considered very \enquote{easy} and not be selected until the very end, at which point the model will predict the label of the duplicate selected first, which will, as labels are conflicting, be wrong, causing accuracy to drop towards the end of the active learning process.

\subsection{Label Smoothing}

A method to obtain soft-labels that does not require any additional information is \textit{label smoothing}~\citep{szegedy2016rethinking, muller2019does}. 
In label smoothing, the original target distribution is mixed with the uniform distribution over all classes: For a given target vector $y$, its smoothed version is calculated as:
$ \mathbf{y}_{smooth} = (1 - \alpha) \cdot \mathbf{y} + \alpha \cdot \frac{1}{K} $, where $K$ is the number of classes and $\alpha$ controls the smoothing strength.
As an example, smoothing the target vector ($A$: 0, $B$: 1, $C$: 0) with $\alpha=0.2$ results in ($A$: 0.06, $B$: 0.86, $C$: 0.06), now a soft-label.
Label smoothing is a form of regulation: intuitively, it dampens the model's prediction confidence, forcing
it to make more \enquote{cautious} predictions. 

We carried out a label-smoothing experiment on the Senti4SD dataset in addition to soft labels.
We train a subset of our Transformer models with hard-labels, soft-labels and different degrees of label-smoothing ($\alpha=0.1$, $0.05$ and $0.03$) respectively, 
and evaluate on all sentiment analysis datasets; refer to Table \ref{tab:smooth-labels} for a comparison of the results. While label smoothing can occasionally improve performance (e.g. on Jira issues), it is more likely to either degrade performance, or not affecting it significantly.

\begin{table}
  \footnotesize
  \begin{center}
    {\tabcolsep=2pt
    \begin{tabular}{llr@{${}\pm{}$}rr}
      \toprule
      \multirow{2}{*}{\textbf{Dataset}} &  \multirow{2}{*}{\textbf{Label Type}} &  \multicolumn{3}{c}{\textbf{F1}}\\
      \cline{3-5}
                                        &                                       &  \multicolumn{2}{c}{$\mu \pm \sigma$} & $max$ \\
      \midrule                                        
      \input{sentiment/soft_label/table}
      \bottomrule
    \end{tabular}
    }
  \end{center}
  \caption{Macro F1 for different label types and datasets: mean, maximum and standard deviation over five runs, each with different seed. All models were trained on the training set of the \stackoverflow{} dataset from \citet{senti4sd}.}
  \label{tab:smooth-labels}
\end{table}

\subsection{Task-Specific Tokens}
\label{subsec:task-tokens}
In the code readability task, our Transformer models cannot compete with manually engineered features used by the baseline.
Since \citet{buse} found that line length is one of the most important features for predicting code readability, we attempt to
provide this information explicitly to our model in form the of special line length tokens, added to line ends.
These tokens range from \texttt{<l1>}, indicating a short line, up to \texttt{<l10>} for very long lines
and are inserted before newline tokens. We fine-tuned the same model with and without these special tokens.

\begin{table}
  \begin{center}
  \begin{tabular}{ll}
   \toprule
   \textbf{Model} & \textbf{Acc.} \\
   \midrule
   \input{readability/results/table}
   \bottomrule
   \end{tabular}
  \end{center}
  \caption{Results for the code readability prediction task. As far as our number are concerned, values are means over five runs with different seeds.}
  \label{tab:readability-results}
\end{table}

Line length tokens failed to improve the performance. In fact, they seem to hurt performance (see Table \ref{tab:readability-results}).
\AtNextCitekey{\defcounter{maxnames}{1}\defcounter{minnames}{1}}
The logistic regression model by \citeauthor{scalabrino2018} outperforms our Transformer model by a wide margin.
We were able to successfully reproduce the results of Scalabrino's model, which was implemented using Weka~\cite{weka}.
We found that for their model, attribute selection is crucial; without it, in our experiments, accuracy dropped significantly (below 60\%).

A simple logistic regression model implemented using \texttt{scikit-learn}~\citep{scikit}, even with attribute selection, was similarly unable
to beat their Weka model; neither was TPOT~\citep{tpot}, a framework for automated machine learning, that automatically
evaluates a large number of combinations of different machine learning algorithms. By first using Weka's attribute selection algorithm 
and feeding selected attribute to a \texttt{scikit-learn} logistic regression model we were eventually able to obtain results close to the Weka-only model.

\subsection{Sentiment Analysis}

Full results, including per-class precision and recall for sentiment classification (Table~\ref{tab:senti-results-expanded}). Finally, as can be seen in table \ref{tab:senti-top-loss}, sentences that are most confusing to the model are hard to classify even for humans. Similar observations can be made in other datasets (e.g., SATD, not shown here).

\begin{table*}
  \footnotesize
  \begin{center}
    {\tabcolsep=4pt%
    \begin{tabular}{@{\extracolsep{2pt}}llcccccccc@{}}
      \toprule
       \multirow{2}{*}{\textbf{Dataset}} &  \multirow{2}{*}{\textbf{Model}} &  \multirow{2}{*}{\textbf{Acc.}} &   \multirow{2}{*}{\textbf{F1}} & \multicolumn{3}{c}{\textbf{Precision}} & \multicolumn{3}{c}{\textbf{Recall}} \\
       \cline{5-7}
       \cline{8-10}
       &   &   &   &  \textbf{Pos.} & \textbf{Neg.} & \textbf{Neu.} & \textbf{Pos.} & \textbf{Neg.} & \textbf{Neu.} \\
       \midrule
       \input{sentiment/results/table}%
       \bottomrule
    \end{tabular} 
    }
  \end{center}
  \caption{Accuracy, macro F1 and per-class precision and recall for different models and datasets. Values reported are means over five runs, each with different seed (only our models). All models were trained on the training set of the \stackoverflow{} sentiment dataset from \citet{senti4sd}.} 
  \label{tab:senti-results-expanded}
\end{table*}

\begin{table*}
   \footnotesize
   \begin{center}
   {\def\arraystretch{1.5}\tabcolsep=3pt
   \begin{tabular}{p{10cm}ccc}
   \dtoprule
   \textbf{Sample}&\textbf{Act. Label}&\textbf{Pred. Label}&\textbf{Agr.}\\
   \dmidrule
   \input{sentiment/top_loss/table}%
   \dbottomrule
   \end{tabular}
   }
   \end{center}
    \caption{Sentences from the test set of the \stackoverflow{}~\citep{senti4sd} with highest loss along with predicted and actual labels and whether all raters agreed on the actual label.}
    \label{tab:senti-top-loss}
 \end{table*}

\section{Further details on Implementation, Run-time, Replication}
\label{appendix:implementation}

\paragraph*{Obtaining back-translation data}

While the original works introducing back-translation used an ad-hoc neural translation model, we found that the most efficient way to obtain back-translations is to load the dataset into Google Sheet and use the \texttt{GOOGLETRANSLATE} macro. An example is available online\footnote{\url{https://docs.google.com/spreadsheets/d/19X8vvV3LF9m2fqUwS9L9yEOPxD-RzshC-fCBTGjkD8I/edit?usp=sharing}}.

\paragraph*{Model implementations}

For all our experiments we use HuggingFace's \texttt{transformers} package~\citep{wolf2020huggingfaces}, a Python library based on
\texttt{pytorch} that implements many different Transformer architectures, including \bert{} and \roberta{}.

\paragraph*{Runtime considerations}
All of our experiments were carried out either on an NVIDIA V100 GPU with 32 GB of memory or on up to three NVIDIA RTX 2080TIs with 10 GB memory each.

Our pre-training regimes are generally affordable even with relatively modest computational budget, although an extensive hyper-parameter search is hardly feasible.  We thus followed common recommendations and only tried a  few parameter combinations.
With a training time of two weeks, pre-training StackOBERTflow \emph{from scratch} was by far the most expensive (especially considering that this was clearly not enough, as it ended up being out-performed by the further pre-trained models).
\emph{Further} pre-training the 12-layer models required considerably less training time, usually below 24 hours (on an NVIDIA V100 GPU).

\paragraph*{Using the pre-trained models}
Our models are publicly available: the StackOBERTflow model can be obtained through the Huggingface Model Hub\footnote{\url{https://huggingface.co/giganticode/StackOBERTflow-comments-small-v1}}; our fine-tuned BERT and RoBERTa models can be downloaded from GitHub\footnote{\url{https://github.com/giganticode/small-datasets-ml-resources/releases/tag/0.1}}.

You can download our pre-trained models and use them for your own experiments.
Our \stackobertflowshort{} model can be automatically downloaded using the 
\texttt{transformers} library. You can instantiate a classification model using 
\texttt{model = AutoModel.from\_pretrained ('giganticode/StackOBERTflow-comments-small-v1')}, and then fine-tune on your task-specific data.  You can also
use our other models: first download the model as ZIP archive from our GitHub page and unpack it; then, likewise,
load them as follows: \texttt{model = AutoModel.from\_pretrained('/path/to/model')}

\paragraph*{Rerunning experiments}
First clone our GitHub repository\footnote{\url{https://github.com/giganticode/small-datasets-ml-resources}}; then run 
\texttt{python -mdl4se.experiments.<experiment>.default --seeds 100 200 300 400 500 --out\_file=result\_file.csv}, where \texttt{experiment} is one of the experiments listed in Table \ref{tab:experiments}. Configuration options and default hyper-parameters can for each experiment
be found in \texttt{/dl4se/config/<experiment>.py}, dataset loading and pre-processing code lies in \texttt{/dl4se/datasets/<experiment>.py}
Note that you cannot use the original datasets, as datasets need to adhere to a specific format. We will provide all of the datasets  upon request.

\begin{table*}
  \begin{center}
  \begin{tabular}{lp{5cm}p{5cm}}
   \toprule
   \textbf{Experiment} & \textbf{Description} & \textbf{Original Dataset URL} \\
   \midrule
\texttt{ar\_miner} & Informative app reviews & \url{https://github.com/jinyyy666/AR_Miner/tree/master/datasets} \\
\texttt{coherence} & Code-comment coherence & \url{http://www2.unibas.it/gscanniello/coherence/} \\
\texttt{comment\_classification} & Comment classification & \url{https://zenodo.org/record/2628361} \\
\texttt{corcod} & Runtime complexity classification & \url{https://github.com/midas-research/corcod-dataset} \\
\texttt{readability} & Code readability classification & \url{https://dibt.unimol.it/report/readability/} \\
\texttt{review\_classification} & Review classification &
\url{https://mast.informatik.uni-hamburg.de/wp-content/uploads/2014/03/REJ_data.zip},
CLAP was requested from the respective authors
\\
\texttt{satd} & Self-admitted debt detection & \url{https://github.com/maldonado/tse.satd.data} \\
\texttt{senti4sd} & Sentiment analysis on Stack Overflow comments and JIRA issues & \url{https://github.com/collab-uniba/Senti4SD}, \url{https://sentidata.github.io/} \\
\texttt{smell\_detection} & Linguistic smell detection & \url{https://github.com/Smfakhoury/SANER-2018-KeepItSimple-} \\
   \end{tabular}
   \end{center}
   
  \caption{Different code modules along with the source of the used datasets.}
  \label{tab:experiments}

\end{table*}

\end{document}

%% file: defs.tex
\def\dssentiforsd{Senti4SD~\cite{senti4sd}}
\def\dsjira{JIRA Issues~\cite{howfarcanwego}}
\def\dsso{StackOverflow~\cite{twodssenti}}
\def\dsapprev{App Reviews~\cite{twodssenti}}

\def\roberta{RoBERTa}
\def\bert{BERT}
\def\stackoverflow{Stack Overflow}
\def\stackobertflow{StackOBERTflow-comments}
\def\stackobertflowshort{StackOBERTflow}
\def\codeberta{CodeBERTa}
\def\codebert{CodeBERT}

%% file: ar_miner/results/table.tex
\stackobertflowshort{} &        90.9 &        89.4 &        88.6 &        85.1 &        88.5 \\
        \bert{} (base) &        90.6 &        88.2 &        89.2 &        85.4 &        88.4 \\
            \bertSOM{} &        92.1 &        90.0 &        91.1 &        87.5 &        90.2 \\
        \bertreviews{} & \bftab 93.3 & \bftab 91.4 & \bftab 91.3 & \bftab 89.9 & \bftab 91.5 \\
   EMNB~\cite{arminer} &        87.7 &        76.1 &        79.7 &        76.4 &        80.0 \\

%% file: smell_detection/results/table.tex
\codeberta{}\tnote{1}               &         81.1 &            81.5 &         81.1 \\
\codebert{}\tnote{1}                &         71.2 &            73.7 &         71.5 \\
SMO Poly\tnote{2}~~\cite{codesmell} &  \bftab 88.8 &     \bftab 91.8 &  \bftab 86.0 \\
SVM RBF\tnote{2}~~\cite{codesmell}  &         74.8 &            76.2 &         73.4 \\
CNN\tnote{2}~~\cite{codesmell}      &         74.5 &            75.6 &         73.5 \\

%% file: corcod/results/table.tex
\codeberta{}                       &         78.2 \\
\codebert{}                        &  \bftab 78.4 \\
Random Forest~\cite{complex}       &         74.3 \\
Logistic Regression~\cite{complex} &         73.2 \\
SVM~\cite{complex}                 &         73.0 \\
SVM+graph2vec~\cite{complex}       &         73.9 \\

%% file: satd/results/table_f1.tex
\textbf{Apache Ant} &          70.2 &                 66.9 &         65.4 &             67.5 &  \bftab 70.3 &                68.8 &         67.2 &             69.0 & 66.0 &  51.2 \\
\textbf{ArgoUML   } &          89.8 &                 90.0 &         89.7 &             89.3 &         89.4 &         \bftab 90.0 &         89.7 &             89.0 & 87.8 &  81.9 \\
\textbf{Columba   } &          90.9 &                 90.4 &         91.0 &             90.0 &         91.4 &                91.4 &  \bftab 91.9 &             90.6 & 85.2 &  75.0 \\
\textbf{EMF       } &          73.5 &                 72.4 &         73.2 &             69.2 &         72.5 &                69.5 &  \bftab 74.4 &             71.0 & 67.9 &  46.2 \\
\textbf{Hibernate } &          88.6 &                 87.5 &         88.2 &             87.4 &         88.7 &                88.3 &  \bftab 88.9 &             87.7 & 82.6 &  76.3 \\
\textbf{JEdit     } &          72.7 &                 70.6 &         71.9 &      \bftab 73.9 &         72.0 &                70.8 &         70.5 &             72.2 & 59.9 &  46.1 \\
\textbf{JFreeChart} &          77.7 &          \bftab 80.7 &         79.2 &             77.6 &         64.1 &                64.7 &         63.8 &             63.0 & 73.9 &  51.3 \\
\textbf{JMeter    } &          87.0 &          \bftab 87.5 &         87.4 &             86.6 &         86.4 &                85.8 &         86.2 &             84.3 & 82.8 &  71.5 \\
\textbf{JRuby     } &          91.1 &                 91.2 &         90.8 &             90.5 &         92.1 &         \bftab 92.4 &         92.3 &             91.4 & 86.3 &  77.3 \\
\textbf{SQuirrel  } &          79.1 &                 78.2 &         78.8 &             78.6 &         80.5 &                79.5 &  \bftab 80.5 &             78.4 & 73.9 &  59.3 \\
\midrule
\textbf{Average   } &   \bftab 82.1 &                 81.5 &         81.6 &             81.1 &         80.7 &                80.1 &         80.5 &             79.7 & 76.6 &  63.6 \\

%% file: comment_classification/results/table.tex
\stackobertflowshort{}                                & 88.4 & 89.6 &  \bftab 90.1 \\
\bertcomments{}                                       & 88.3 &  \bftab 90.6 & 89.0 \\
\bert{}                                               &  \bftab 88.4 & 90.5 & 89.1 \\
Naive Bayes Multinomial~\cite{comment-classification} & 84.3 & 82.0 & 87.2 \\

%% file: sentiment/results/condensed.tex
\multirow{11}{*}{\makecell{\dssentiforsd{}\\(Test Set)}} & BERT (base) &           87.9 &           87.8   \\
                                       & \bertSOML{} &  {\bftab 89.4} &  {\bftab 89.4}   \\
                                       & \bertSOMM{} &           88.8 &           88.7   \\
                                       & \bertSOM{} &           88.8 &           88.7   \\
                                       & DistilRoBERTa (small) &           87.4 &           87.3   \\
                                       & RoBERTa (small, no pretr.) &           78.4 &           77.8   \\
                                       & Senti4SD~\cite{senti4sd} &              - &           86.0   \\
                                       & SentiCR~\cite{senti4sd} &              - &           82.0   \\
                                       & SentiStrengthSE~\cite{senti4sd} &              - &           80.0  \\
                                       & SentiStrength~\cite{senti4sd} &              - &           84.0   \\
                                       & StackOBERTflow &           88.7 &           88.6   \\
\cline{1-4}
\multirow{12}{*}{\dsapprev{}} & BERT (base) &           64.9 &           50.2  \\
                                       & \bertSOML{} &           66.0 &           51.9  \\
                                       & \bertSOMM{} &           67.7 &           53.1  \\
                                       & \bertSOM{} &           68.3 &           53.7   \\
                                       & DistilRoBERTa (small) &           60.9 &           48.5   \\
                                       & NLTK~\cite{howfarcanwego} &           54.0 &           40.8   \\
                                       & RoBERTa (small, no pretr.) &           48.0 &           42.2  \\
                                       & SentiStrength+AC0-SE~\cite{howfarcanwego} &           58.9 &           46.6  \\
                                       & SentiStrength~\cite{howfarcanwego} &           62.5 &           48.2   \\
                                       & StackOBERTflow &  {\bftab 72.0} &  {\bftab 57.4}   \\
                                       & Stanford CoreNLP SO~\cite{howfarcanwego} &           41.6 &           35.5   \\
                                       & Stanford CoreNLP~\cite{howfarcanwego} &           69.5 &           56.0   \\
\cline{1-4}
\multirow{12}{*}{\dsjira{}} & BERT (base) &  {\bftab 95.7} &  {\bftab 95.0}  \\
                                       & \bertSOML{} &           94.8 &           93.8  \\
                                       & \bertSOMM{} &           95.2 &           94.5   \\
                                       & \bertSOM{} &           95.1 &           94.2   \\
                                       & DistilRoBERTa (small) &           93.7 &           92.7   \\
                                       & NLTK~\cite{howfarcanwego} &           29.8 &           46.5   \\
                                       & RoBERTa (small, no pretr.) &           84.0 &           80.1   \\
                                       & SentiStrength+AC0-SE~\cite{howfarcanwego} &           76.0 &           87.0   \\
                                       & SentiStrength~\cite{howfarcanwego} &           77.1 &           85.4  \\
                                       & StackOBERTflow &           93.5 &           92.5  \\
                                       & Stanford CoreNLP SO~\cite{howfarcanwego} &           36.0 &           44.2   \\
                                       & Stanford CoreNLP~\cite{howfarcanwego} &           67.6 &           73.7   \\
\cline{1-4}
\multirow{12}{*}{\dsso{}} & BERT (base) &           79.9 &           47.6   \\
                                       & \bertSOML{} &           79.2 &           43.3   \\
                                       & \bertSOMM{} &  {\bftab 80.3} &           48.6   \\
                                       & \bertSOM{} &           80.1 &           47.9   \\
                                       & DistilRoBERTa (small) &           78.8 &           44.3   \\
                                       & NLTK~\cite{howfarcanwego} &           77.9 &           43.2   \\
                                       & RoBERTa (small, no pretr.) &           75.5 &           42.2  \\
                                       & SentiStrength+AC0-SE~\cite{howfarcanwego} &           78.0 &           46.8  \\
                                       & SentiStrength~\cite{howfarcanwego} &           69.5 &  {\bftab 49.5}  \\
                                       & StackOBERTflow &           79.3 &           44.1   \\
                                       & Stanford CoreNLP SO~\cite{howfarcanwego} &           75.9 &           47.5  \\
                                       & Stanford CoreNLP~\cite{howfarcanwego} &           40.3 &           35.5   \\

%% file: sentiment/soft_label/condensedtable.tex
\multirow{2}{*}{\dsapprev{}} & All Label Votes &  {\bftab 57.5} &  2.2 &           59.1 \\
                                       & Majority Label &           57.4 &  1.6 &           59.7 \\
\cline{1-5}
\multirow{2}{*}{\dsjira{}} & All Label Votes &           93.5 &  1.0 &  {\bftab 94.7} \\
                                       & Majority Label &           92.5 &  0.5 &           93.2 \\
\cline{1-5}
\multirow{2}{*}{\dsso{}} & All Label Votes &           42.4 &  1.0 &           43.8 \\
                                       & Majority Label &  {\bftab 44.1} &  2.9 &           46.4 \\
\cline{1-5}
\multirow{2}{*}{\makecell{\dssentiforsd{}}} & All Label Votes  &  {\bftab 89.1} &  0.5 &           89.6 \\
                                       & Majority Label &           88.6 &  0.5 &           89.1 \\

%% file: review_classification/results/table.tex
\stackobertflowshort{}                               & 61.1 & 42.0 & 79.4 & 49.8 & 58.1 \\
\stackobertflowshort{}+BT                            & 62.5 & 43.3 & 79.0 & 50.2 & 58.7 \\
\stackobertflowshort{}+BTT                           & 62.7 & 44.3 & 79.6 & 50.6 & 59.3 \\
\bert{} (base)                                       & 61.8 & 39.7 & 79.4 & 49.4 & 57.6 \\
\bert{} (base)+BT                                    & 62.0 & 43.1 & 79.4 & 50.2 & 58.7 \\
\bert{} (base)+BTT                                   & 62.5 & 42.6 & 80.0 & 50.1 & 58.8 \\
\bertreviews{}                                       &  \bftab 64.1 & 44.7 &  \bftab 80.2 & 51.0 & 60.0 \\
\bertreviews{}+BT                                    & 63.2 & 43.9 & 79.5 & 51.0 & 59.4 \\
\bertreviews{}+BTT                                   & 63.7 & 46.7 & 80.1 & 51.2 &  \bftab 60.4 \\
Decision Tree\tnote{1}~~\cite{review-classification} & 62.0 & 42.0 & 54.0 & 50.0 & 52.0 \\
Naive Bayes\tnote{1}~~\cite{review-classification}   & 62.0 &  \bftab 47.0 & 54.0 &  \bftab 53.0 & 54.0 \\

%% file: backtrans/table_clap.tex
\multirow{4}{*}{CLAP} & Train+Test &  \bftab 89.2 &  \bftab 81.4 &     \bftab 84.2 &  \bftab 82.3 \\
     & None &         88.1 &         79.7 &            80.6 &         81.7 \\
     & Test &         88.3 &         80.2 &            81.8 &         81.5 \\
     & Train &         88.9 &         81.1 &            83.7 &         82.0 \\

%% file: backtrans/table_senti4sd.tex
\multirow{4}{*}{Senti4SD} & Train+Test &  \bftab 88.7 &  \bftab 88.6 &     \bftab 88.7 &  \bftab 88.5 \\
         & None &         88.4 &         88.2 &            88.2 &         88.4 \\
         & Test &         88.2 &         88.0 &            88.1 &         88.0 \\
         & Train &         88.4 &         88.3 &            88.3 &         88.5 \\

%% file: coherence/results/table.tex
\codeberta{}                    &         81.4 &         80.1 \\
\codeberta{}+ITT                &         82.5 &         80.9 \\
\codebert{}                     &         81.6 &         80.0 \\
SVM~\cite{corazza}\tnote{1}     &  \bftab 83.5 &         81.3 \\
SVM~\cite{corazza}\tnote{2}     &         83.3 &  \bftab 82.2 \\
SVM+Word~Emb.~\cite{corazza-we} &         80.5 &            - \\

%% file: samplestable.tex
\begin{table*}
  \footnotesize
  \makegapedcells
  \begin{center}
  {\def\arraystretch{1}\tabcolsep=2pt\setcellgapes{3pt}
  \begin{tabular}{clc}
    \toprule
    \textbf{Name}&\textbf{Sample}&\textbf{Class}\\
    \midrule
    \makecell[c]{Sentiment Classification\\(Stack Overflow)~\cite{senti4sd}} &
    I want them to resize based on the length of the data they're showing. &
    neutral \\

    \makecell[c]{Sentiment Classification\\(Stack Overflow)~\cite{twodssenti}} &
    When I run my client, it throws the following exception. &
    negative \\

    \makecell[c]{Sentiment Classification\\(JIRA Issues)~\cite{howfarcanwego}} &  
    This is always a really bad way to design software. &
    negative \\
    
    \makecell[c]{Sentiment Classification\\(App Reviews)~\cite{twodssenti}} &
    amazing! a must have app &
    positive \\

    \makecell[c]{Informative App Review\\Detection~\cite{arminer}} & 
    not able to download any pictures please fix these bugs immediately &
    informative \\
    
    \makecell[c]{App Review\\Classification~\cite{review-classification}} & 
    Best game I've played on Android &
    rating \\
    
    \makecell[c]{App Review\\Classification~\cite{clap}} &
    good but... it has ads...please remove ads from this... & 
    usability \\
    
    \midrule
    
    \makecell[c]{Self-Admitted Technical\\Debt Detection~\cite{satd-orig}} &
    \texttt{// FIXME: Is "No Namespace is Empty Namespace" really OK?} &
    SATD \\
    
    \makecell[c]{Comment Classification\\\cite{comment-classification}} &
    \texttt{@return a string for throwing} &
    usage \\
    
    \midrule    
    
    \makecell[c]{Code-Comment Coherence\\Prediction~\cite{corazza-ds}} & 
    \begin{minipage}{3in}
    \begin{verbatim}
/**
  * Returns the current number of milk units in
  * the inventory.
  * @return int
  */
  public int getMilk() {
      return milk;
  }
     \end{verbatim}
     \end{minipage} &
     coherent \\
    
    \makecell[c]{Linguistic Smell\\Detection~\cite{codesmell}} & 
    \begin{minipage}{6cm}
    \begin{verbatim}
public void ToSource(StringBuilder sb) {
    sb.append(";");
    this.NewLine(sb);
}
    \end{verbatim}
    \end{minipage} &
    \makecell[c]{smelly\\(transform method\\does not return)} \\

    \makecell[c]{Code Runtime Complexity\\Classification~\cite{complex}} & 
    \begin{minipage}{6cm}

    \begin{verbatim}    
class GFG {
  static int minJumps(int arr[], int n) {
    int[] jumps = new int[n];
    int min;
    jumps[n - 1] = 0;
    for (int i = n - 2; i >= 0; i--) {
      if (arr[i] == 0) jumps[i] = Integer.MAX_VALUE;
      else if (arr[i] >= n - i - 1) jumps[i] = 1;
      else { ... }
    }
    return jumps[0];
  }
  public static void main(String[] args) {...}
}
    \end{verbatim}
    \end{minipage} &
    $\mathcal{O}(n\log{}n)$ \\

    \makecell[c]{Code Readability\\Prediction~\cite{scalabrino2016}} &
\begin{minipage}{4cm}
\begin{verbatim}    
    
@Override
public void configure(Configuration cfg) {
  super.configure(cfg);
  cfg.setProperty(Environment.USE_SECOND_LEVEL_CACHE, ...);
  cfg.setProperty(Environment.GENERATE_STATISTICS, ...);
  cfg.setProperty(Environment.USE_QUERY_CACHE, "false" );
  ... // more cfg.setProperty calls
}    
    \end{verbatim}
    \end{minipage} &
    readable \\    

    \bottomrule
  \end{tabular}
  }
  \end{center}
  \caption{Arbitrarily selected examples from the datasets along with their class. Some of the code samples have been shortened with ellipses.} 
  \label{tab:ds-examples}
\end{table*}

%% file: satd/dataset/table.tex
\begin{tabular}{rrrr}
\toprule
\textbf{{}}&\textbf{SATD}&\textbf{not SATD}&\textbf{Total}\\
\midrule
\textbf{Apache Ant} &  \numprint{131} &  \numprint{3967} &  \numprint{4098} \\
\textbf{ArgoUML   } & \numprint{1413} &  \numprint{8039} &  \numprint{9452} \\
\textbf{Columba   } &  \numprint{204} &  \numprint{6264} &  \numprint{6468} \\
\textbf{EMF       } &  \numprint{104} &  \numprint{4286} &  \numprint{4390} \\
\textbf{Hibernate } &  \numprint{472} &  \numprint{2496} &  \numprint{2968} \\
\textbf{JEdit     } &  \numprint{256} & \numprint{10066} & \numprint{10322} \\
\textbf{JFreeChart} &  \numprint{209} &  \numprint{4199} &  \numprint{4408} \\
\textbf{JMeter    } &  \numprint{374} &  \numprint{7683} &  \numprint{8057} \\
\textbf{JRuby     } &  \numprint{622} &  \numprint{4275} &  \numprint{4897} \\
\textbf{SQuirrel  } &  \numprint{286} &  \numprint{6929} &  \numprint{7215} \\
\midrule
\textbf{Total     } & \numprint{4071} & \numprint{58204} & \numprint{62275} \\
\bottomrule
\end{tabular}

%% file: sentiment/soft_label/table.tex
\multirow{8}{*}{\dsapprev{}} & All Label Votes ($\alpha$ = 0.03) &           56.9 &  3.3 &  {\bftab 61.6} \\
                                       & All Label Votes ($\alpha$ = 0.05) &           56.2 &  2.9 &           58.8 \\
                                       & All Label Votes ($\alpha$ = 0.1) &           56.4 &  2.6 &           59.0 \\
                                       & All Label Votes (no smooth.) &  {\bftab 57.5} &  2.2 &           59.1 \\
                                       & Majority Label ($\alpha$ = 0.03) &           55.7 &  3.8 &           60.4 \\
                                       & Majority Label ($\alpha$ = 0.05) &           55.9 &  1.5 &           57.6 \\
                                       & Majority Label ($\alpha$ = 0.1) &           56.7 &  1.6 &           58.8 \\
                                       & Majority Label (no smooth.) &           57.4 &  1.6 &           59.7 \\
\cline{1-5}
\multirow{8}{*}{\dsjira{}} & All Label Votes ($\alpha$ = 0.03) &           92.7 &  1.4 &           94.2 \\
                                       & All Label Votes ($\alpha$ = 0.05) &           93.2 &  0.7 &           94.0 \\
                                       & All Label Votes ($\alpha$ = 0.1) &           93.1 &  1.1 &           94.2 \\
                                       & All Label Votes (no smooth.) &           93.5 &  1.0 &  {\bftab 94.7} \\
                                       & Majority Label ($\alpha$ = 0.03) &           91.6 &  1.9 &           94.3 \\
                                       & Majority Label ($\alpha$ = 0.05) &           92.4 &  1.3 &           93.6 \\
                                       & Majority Label ($\alpha$ = 0.1) &  {\bftab 93.5} &  0.7 &           94.3 \\
                                       & Majority Label (no smooth.) &           92.5 &  0.5 &           93.2 \\
\cline{1-5}
\multirow{8}{*}{\dsso{}} & All Label Votes ($\alpha$ = 0.03) &           41.6 &  1.3 &           43.6 \\
                                       & All Label Votes ($\alpha$ = 0.05) &           42.5 &  2.4 &           46.1 \\
                                       & All Label Votes ($\alpha$ = 0.1) &           41.0 &  1.4 &           42.7 \\
                                       & All Label Votes (no smooth.) &           42.4 &  1.0 &           43.8 \\
                                       & Majority Label ($\alpha$ = 0.03) &           42.8 &  3.5 &  {\bftab 47.4} \\
                                       & Majority Label ($\alpha$ = 0.05) &           42.7 &  2.3 &           45.6 \\
                                       & Majority Label ($\alpha$ = 0.1) &           43.1 &  1.4 &           45.2 \\
                                       & Majority Label (no smooth.) &  {\bftab 44.1} &  2.9 &           46.4 \\
\cline{1-5}
\multirow{8}{*}{\makecell{\dssentiforsd{}\\(Test Set)}} & All Label Votes ($\alpha$ = 0.03) &           88.9 &  0.2 &           89.2 \\
                                       & All Label Votes ($\alpha$ = 0.05) &           89.0 &  0.6 &  {\bftab 89.9} \\
                                       & All Label Votes ($\alpha$ = 0.1) &           89.0 &  0.4 &           89.5 \\
                                       & All Label Votes (no smooth.) &  {\bftab 89.1} &  0.5 &           89.6 \\
                                       & Majority Label ($\alpha$ = 0.03) &           88.4 &  0.5 &           89.0 \\
                                       & Majority Label ($\alpha$ = 0.05) &           88.4 &  0.4 &           89.1 \\
                                       & Majority Label ($\alpha$ = 0.1) &           88.5 &  0.3 &           88.8 \\
                                       & Majority Label (no smooth.) &           88.6 &  0.5 &           89.1 \\

%% file: readability/results/table.tex
CodeBERTa                                 &         73.1 \\
CodeBERTa+LLT                             &         72.5 \\
CodeBERT                                  &         69.3 \\
Logistic Regression~\cite{scalabrino2018} &  \bftab 84.0 \\

%% file: sentiment/results/table.tex
\multirow{12}{*}{\dsapprev{}} & BERT (base) &           64.9 &           50.2 &           76.1 &            84.7 &            9.8 &           86.6 &           41.4 &           25.6 \\
                                       & \bertSOML{} &           66.0 &           51.9 &           77.2 &            83.7 &           10.6 &           84.7 &           46.8 &           27.2 \\
                                       & \bertSOMM{} &           67.7 &           53.1 &           77.2 &            85.4 &           11.7 &           87.4 &           47.4 &           27.2 \\
                                       & \bertSOM{} &           68.3 &           53.7 &           77.2 &            86.9 &           12.7 &           88.0 &           47.8 &           28.8 \\
                                       & DistilRoBERTa (small) &           60.9 &           48.5 &           75.3 &            74.7 &           10.1 &           77.3 &           43.5 &           28.8 \\
                                       & NLTK~\cite{howfarcanwego} &           54.0 &           40.8 &           75.1 &  {\bftab 100.0} &            9.3 &           81.2 &           16.9 &  {\bftab 44.0} \\
                                       & RoBERTa (small, no pretr.) &           48.0 &           42.2 &           73.8 &            68.9 &            8.5 &           53.0 &           42.0 &           42.4 \\
                                       & SentiStrength+AC0-SE~\cite{howfarcanwego} &           58.9 &           46.6 &           74.1 &            92.9 &           10.6 &           81.7 &           30.0 &           40.0 \\
                                       & SentiStrength~\cite{howfarcanwego} &           62.5 &           48.2 &           74.5 &            81.5 &           11.3 &           86.6 &           33.8 &           32.0 \\
                                       & StackOBERTflow &  {\bftab 72.0} &  {\bftab 57.4} &           79.7 &            86.4 &           15.1 &  {\bftab 88.4} &           56.6 &           29.6 \\
                                       & Stanford CoreNLP SO~\cite{howfarcanwego} &           41.6 &           35.5 &           77.0 &            47.0 &            8.4 &           25.3 &           66.9 &           32.0 \\
                                       & Stanford CoreNLP~\cite{howfarcanwego} &           69.5 &           56.0 &  {\bftab 83.1} &            66.7 &  {\bftab 17.6} &           71.5 &  {\bftab 75.4} &           24.0 \\
\cline{1-10}
\multirow{12}{*}{\dsjira{}} & BERT (base) &  {\bftab 95.7} &  {\bftab 95.0} &           92.4 &            97.2 &              - &  {\bftab 94.0} &           96.4 &              - \\
                                       & \bertSOML{} &           94.8 &           93.8 &           93.3 &            95.4 &              - &           89.7 &  {\bftab 97.0} &              - \\
                                       & \bertSOMM{} &           95.2 &           94.5 &           92.2 &            96.8 &              - &           92.9 &           96.3 &              - \\
                                       & \bertSOM{} &           95.1 &           94.2 &           93.1 &            96.0 &              - &           91.0 &           96.9 &              - \\
                                       & DistilRoBERTa (small) &           93.7 &           92.7 &           89.9 &            95.5 &              - &           90.1 &           95.3 &              - \\
                                       & NLTK~\cite{howfarcanwego} &           29.8 &           46.5 &           84.0 &  {\bftab 100.0} &              - &           36.2 &           26.9 &              - \\
                                       & RoBERTa (small, no pretr.) &           84.0 &           80.1 &           78.6 &            86.3 &              - &           67.0 &           91.7 &              - \\
                                       & SentiStrength+AC0-SE~\cite{howfarcanwego} &           76.0 &           87.0 &  {\bftab 94.8} &            99.6 &              - &           88.3 &           70.4 &              - \\
                                       & SentiStrength~\cite{howfarcanwego} &           77.1 &           85.4 &           85.0 &            99.3 &              - &           92.1 &           70.3 &              - \\
                                       & StackOBERTflow &           93.5 &           92.5 &           87.8 &            96.4 &              - &           92.1 &           94.1 &              - \\
                                       & Stanford CoreNLP SO~\cite{howfarcanwego} &           36.0 &           44.2 &           63.5 &            72.4 &              - &           25.2 &           40.9 &              - \\
                                       & Stanford CoreNLP~\cite{howfarcanwego} &           67.6 &           73.7 &           72.6 &            94.5 &              - &           62.1 &           70.1 &              - \\
\cline{1-10}
\multirow{12}{*}{\dsso{}} & BERT (base) &           79.9 &           47.6 &           32.1 &            79.9 &           82.0 &           15.9 &           21.1 &           95.7 \\
                                       & \bertSOML{} &           79.2 &           43.3 &           29.8 &            78.2 &           81.2 &           13.4 &           13.8 &  {\bftab 96.2} \\
                                       & \bertSOMM{} &  {\bftab 80.3} &           48.6 &           33.7 &            79.8 &           82.4 &           16.2 &           23.0 &           95.9 \\
                                       & \bertSOM{} &           80.1 &           47.9 &  {\bftab 34.9} &   {\bftab 80.0} &           82.1 &           16.8 &           20.8 &           95.9 \\
                                       & DistilRoBERTa (small) &           78.8 &           44.3 &           31.5 &            61.3 &           81.5 &           14.8 &           16.2 &           95.2 \\
                                       & NLTK~\cite{howfarcanwego} &           77.9 &           43.2 &           31.7 &            62.5 &           81.5 &           24.4 &            8.4 &           94.1 \\
                                       & RoBERTa (small, no pretr.) &           75.5 &           42.2 &           21.1 &            37.2 &           81.6 &           10.8 &           21.1 &           90.7 \\
                                       & SentiStrength+AC0-SE~\cite{howfarcanwego} &           78.0 &           46.8 &           31.2 &            50.0 &           82.6 &           22.1 &           18.5 &           93.0 \\
                                       & SentiStrength~\cite{howfarcanwego} &           69.5 &  {\bftab 49.5} &           20.0 &            39.7 &           85.8 &  {\bftab 35.9} &           43.3 &           77.2 \\
                                       & StackOBERTflow &           79.3 &           44.1 &           32.1 &            78.1 &           81.5 &           15.7 &           13.7 &           96.1 \\
                                       & Stanford CoreNLP SO~\cite{howfarcanwego} &           75.9 &           47.5 &           31.7 &            36.5 &           83.6 &           14.5 &           36.5 &           88.6 \\
                                       & Stanford CoreNLP~\cite{howfarcanwego} &           40.3 &           35.5 &           23.1 &            17.7 &  {\bftab 88.4} &           34.4 &  {\bftab 83.7} &           34.4 \\
\cline{1-10}
\multirow{11}{*}{\makecell{\dssentiforsd{}\\(Test Set)}} & BERT (base) &           87.9 &           87.8 &           92.8 &            88.0 &           83.7 &           93.0 &           83.1 &           86.8 \\
                                       & \bertSOML{} &  {\bftab 89.4} &  {\bftab 89.4} &  {\bftab 93.4} &   {\bftab 88.2} &           86.8 &           93.6 &           86.8 &  {\bftab 87.4} \\
                                       & \bertSOMM{} &           88.8 &           88.7 &           93.2 &            85.6 &           87.1 &  {\bftab 94.9} &           86.7 &           84.7 \\
                                       & \bertSOM{} &           88.8 &           88.7 &           92.7 &            86.9 &           86.7 &           94.5 &           85.9 &           85.7 \\
                                       & DistilRoBERTa (small) &           87.4 &           87.3 &           92.3 &            81.2 &           87.9 &           92.7 &           88.6 &           81.8 \\
                                       & RoBERTa (small, no pretr.) &           78.4 &           77.8 &           84.0 &            72.4 &           77.6 &           86.1 &           69.1 &           78.1 \\
                                       & Senti4SD~\cite{senti4sd} &              - &           86.0 &           92.0 &            80.0 &           87.0 &           92.0 &           89.0 &           80.0 \\
                                       & SentiCR~\cite{senti4sd} &              - &           82.0 &           88.0 &            79.0 &           79.0 &           90.0 &           73.0 &           82.0 \\
                                       & SentiStrengthSE~\cite{senti4sd} &              - &           80.0 &           89.0 &            75.0 &           75.0 &           83.0 &           79.0 &           77.0 \\
                                       & SentiStrength~\cite{senti4sd} &              - &           84.0 &           89.0 &            67.0 &  {\bftab 95.0} &           92.0 &  {\bftab 96.0} &           64.0 \\
                                       & StackOBERTflow &           88.7 &           88.6 &           93.0 &            85.6 &           86.9 &           94.4 &           87.0 &           84.6 \\

%% file: sentiment/top_loss/table.tex
 This worked for me. &  positive &  neutral &  no \\
\hline
 In addition to firebug (which should be your first port of call), the will also tell you where a given style is sourced from, just in case IE {-} shock, horror {-} should be different. &  negative &  neutral &  no \\
\hline
 The named scopes already proposed are pretty fine. The clasic way to do it would be: &  positive &  neutral &  yes \\
\hline
 I have a basic app written with ATL, using the wizard with VS2008. I have a treeview in the left side of the app. I see how to (painfully) add tree items. Question is how do I show a menu when the mouse is right clicked? How do I trap any click events on each item that could be selected? &  negative &  neutral &  yes \\
\hline
 Your implementation looks absolutely fine to me! A range{-}based away subscript is a type for performance reasons. It does not copy the indicated sub{-}array, instead it just points to the range defined by the you provide to the subscript. &  positive &  neutral &  no \\
\hline
 Is it possible (or desirable?!) to set up to behave more like ? For example, instead of writing why can't I just write Similarly, instead of why not just &  negative &  neutral &  no \\
\hline
 This whole DB is almost entirely read only so I'm not too worried about it changing. &  positive &  neutral &  no \\
\hline
 I would agree except they are related and I really hated the idea of writing 4 separate questions since they seemed to close. &  negative &  positive &  yes \\
\hline
 I understand its not a desirable circumstance, however if I NEEDED to have some kind of HTML within JSON tags, e.g.: is this possible to do in Python without requiring to to be escaped beforehand? It will be a string initially so I was thinking about writing a regular expression to attempt to match and escape these prior to processing, but I just want to make sure there isn't an easier way. &  negative &  neutral &  no \\
\hline
 You can easily define a comparator for a one{-}level , so that lookup becomes way less cumbersome. There is no reason of being afraid of that. The comparator defines an ordering of the \_Key template argument of the map. It can then also be used for the multimap and set collections. An example: &  positive &  neutral &  no \\
\hline
 I'm new to Flash but want to create a nice video for a product. It takes a long time to make a nice looking presentation , and I'm hoping for a jump start. Are there any good templates which are free on the internet where I can quickly change the text in ,for example, to make my video? I've tried looking in google, and there are too many websites, many of which look gimmicky. Any recommendations? (A video like this one {-} {-} would be amazing!) &  neutral &  positive &  no \\